\documentclass[aps, prd, onecolumn, tightenlines, notitlepage, superscriptaddress, nofootinbib, preprintnumbers, floatfix, showkeys,11pt]{revtex4-2}

\usepackage[normalem]{ulem}
\usepackage{amssymb}
\usepackage{amsmath}
\usepackage{graphicx}
\usepackage{url}
\usepackage{ulem}
\usepackage[utf8]{inputenc}
\usepackage{slashed}
\usepackage{lipsum}
\usepackage[T1]{fontenc}
\usepackage{appendix}
\usepackage{csquotes}
\usepackage[x11names,table]{xcolor}
\usepackage{comment}
\usepackage[export]{adjustbox}
\usepackage{textgreek}
\usepackage{caption, subcaption}
\usepackage{ragged2e} 
\DeclareCaptionJustification{justified}{\justifying}
\usepackage{booktabs}
\usepackage{orcidlink}
\usepackage{hhline}

\usepackage{hyperref}

\hypersetup{colorlinks,citecolor= nicered,linkcolor= blue}
\definecolor{nicered}{rgb}{0.7,0.1,0.1}
\definecolor{nicegreen}{rgb}{0.1,0.5,0.1}

\def\cevns{CE\textnu NS}
\def\eves{E\textnu ES}

\def\d{\mathrm{d}}
\newcommand{\qtransfer}{\left|\mathbf{q}\right|}

\definecolor{amber}{rgb}{1.0, 0.49, 0.0}
\definecolor{yellow-green}{rgb}{0.6, 0.8, 0.2}
\definecolor{gray(x11gray)}{rgb}{0.75, 0.75, 0.75}
\definecolor{lightsalmon}{rgb}{1.0, 0.63, 0.48}
\definecolor{myblue}{cmyk}{0.65, 0.37, 0.0, 0.19}
\definecolor{darkraspberry}{rgb}{0.53, 0.15, 0.34}

\usepackage{multirow}

\AtBeginDocument{\hypersetup{citecolor=darkraspberry,linkcolor=darkraspberry,urlcolor=darkraspberry}}

\begin{document}

\title{{\Large Neutrino electromagnetic properties after the first\\ $pp$ solar neutrino observation at XENONnT}}

\author{Valentina De Romeri~\orcidlink{0000-0003-3585-7437}}\email{deromeri@ific.uv.es}
\affiliation{Instituto de F\'{i}sica Corpuscular (IFIC), CSIC‐Universitat de Val\`encia, E-46980 Valencia, Spain}

\author{Dimitrios K. Papoulias~\orcidlink{0000-0003-0453-8492}}\email{dimitrios.papoulias@uni-hamburg.de}
\affiliation{Institute of Experimental Physics, University of Hamburg, 22761, Hamburg, Germany}

\author{Federica Pompa~\orcidlink{0000-0002-9591-8361}}
\email{federica.pompa@lngs.infn.it}
\affiliation{Istituto Nazionale di Fisica Nucleare (INFN), Laboratori Nazionali del Gran Sasso, 67100 Assergi, L’Aquila (AQ), Italy
}

\author{Christoph A. Ternes~\orcidlink{0000-0002-7190-1581}}
\email{christoph.ternes@lngs.infn.it}
\affiliation{Istituto Nazionale di Fisica Nucleare (INFN), Laboratori Nazionali del Gran Sasso, 67100 Assergi, L’Aquila (AQ), Italy
}
\affiliation{Gran Sasso Science Institute, Viale F. Crispi 7, L’Aquila, 67100, Italy}

\keywords{dark matter detectors, \eves, $pp$ solar neutrinos, electromagnetic properties, sterile dipole}

\begin{abstract}
The XENONnT Collaboration has recently reported the first measurement of low-energy solar neutrinos through elastic neutrino-electron scattering in a dark matter detector, opening a new window on Standard Model predictions and on neutrino electromagnetic properties. We analyze these data to extract the $pp$ neutrino flux normalization and the weak mixing angle at low momentum transfer ($\sim 0.1$ MeV). We also derive bounds on the neutrino charge radii, magnetic moments, and millicharges, as well as on the sterile-neutrino dipole portal. Since the $pp$-flux determination displays a mild excess, we further investigate the role of the $^{14}$C background in the fit and assess how it affects the extracted parameters. Our results show that dark matter detectors now offer a probe of neutrino physics that is complementary to dedicated precision experiments.
\end{abstract}

\maketitle

\section{Introduction}

Direct-detection experiments have driven the search for dark matter (DM) over the past few decades~\cite{Billard:2021uyg,Schumann:2019eaa}, largely inspired by the weakly interacting massive particle (WIMP) paradigm. This strategy has traditionally focused on nuclear recoils induced by DM-nucleus scattering, but the physics reach of these experiments extends well beyond their primary target. Neutrinos, in particular, have long been recognized as an unavoidable component of the direct-detection program, since astrophysical neutrinos like those produced in the Sun can produce measurable nuclear recoils in low-background detectors~\cite{Monroe:2007xp,Vergados:2008jp,Strigari:2009bq,Billard:2013qya}. As detector masses have grown and backgrounds have fallen, neutrino interactions have evolved from an irreducible background into a physics signal on their own.

The advent of large-scale liquid-xenon time-projection chambers (LXe TPCs) has accelerated this shift. Experiments such as XENONnT~\cite{XENON:2023cxc}, LUX-ZEPLIN (LZ)~\cite{LZ:2022lsv}, and PandaX-4T~\cite{PandaX:2024qfu} have reached exposures and background levels sufficient to observe solar-neutrino interactions through complementary channels. One of them is coherent elastic neutrino-nucleus scattering (\cevns)~\cite{Freedman:1973yd, Abdullah:2022zue}. \cevns~induced by $^8$B solar neutrinos has now been reported by XENONnT~\cite{XENON:2024ijk,XENONCollaboration:2026xfz}, PandaX-4T~\cite{PandaX:2024muv}, and LZ~\cite{LZ:2025igz}. Because \cevns~is approximately flavor-independent, it probes the total solar-neutrino flux with a particularly clean sensitivity to neutrino-quark interactions. Given current nuclear recoil-energy thresholds, only the $^8$B component is presently accessible this way (and, with much smaller contribution, the $hep$ component); the more abundant $pp$ and other lower-energy components (like $^7$Be) would require substantially lower thresholds.

Solar neutrinos can also be detected through elastic neutrino-electron scattering (\eves), producing electronic instead of nuclear recoils. Unlike \cevns, \eves~is flavor-sensitive, since electron- and non-electron-neutrino cross sections differ. Consequently, the measured rate depends not only on the solar neutrino flux but also on the electron-neutrino survival probability. Electron recoils therefore provide information complementary to that obtained from \cevns, combining sensitivity to solar neutrino oscillations with probes of neutrino interactions with electrons. The implications of DM direct-detection electron-recoil data for neutrino physics and new physics more broadly have been extensively studied in the literature, though until now such analyses relied on lower-statistics datasets that yielded flux upper limits rather than actual measurements, or provided projected sensitivities, see, e.g., Refs.~\cite{Harnik:2012ni,Cerdeno:2016sfi,Essig:2018tss,DARWIN:2020bnc,Amaral:2020tga,AristizabalSierra:2020edu,deGouvea:2021ymm,AristizabalSierra:2021fuc,Majumdar:2021vdw,AtzoriCorona:2022jeb,Giunti:2023yha,Maity:2024aji,DeRomeri:2024dbv,AtzoriCorona:2025xwr}. 

However, the low thresholds now achievable in LXe TPCs mark a turning point for this program. XENONnT has recently reported the first detection of $pp$ solar neutrinos via electron recoils~\cite{XENON:2026rbz}, a result made possible by an electron-recoil threshold of 1~keV, which makes the detector sensitive to neutrino energies down to $\sim$17~keV, together with improved rejection of the $^{222}$Rn/$^{214}$Pb background. 
This result underscores the value of lowering recoil thresholds over simply scaling up detector volume, and demonstrates that modern LXe DM detectors have evolved into competitive neutrino observatories, capable of accessing solar-neutrino components that were previously measured only by dedicated neutrino experiments. In particular, it complements earlier determinations of the $pp$ flux by radiochemical gallium experiments (GALLEX/GNO~\cite{Kaether:2010ag} and SAGE~\cite{SAGE:1999nng}), which measured the integrated capture rate above 233~keV, and by Borexino~\cite{BOREXINO:2018ohr}, which
performed the first real-time spectral detection of $pp$ neutrinos in a $\sim$100-ton liquid-scintillator detector. A consistent, yet less statistically significant, \eves~signal previously reported by PandaX-4T~\cite{PandaX:2024jjs,PandaX:2026qdp} further reinforces this result, marking an important extension of the neutrino program at DM detectors to lower energies.

This first observation of $pp$ solar neutrinos via electron recoils at such low energies opens several avenues beyond a flux measurement: it allows for a determination of the weak mixing angle at lower momentum transfer than \cevns, probes of ultra-small quasi-Dirac mass splittings~\cite{Leone:2026knr}, non-standard interactions~\cite{Amaral:2026vkt}, and it is also sensitive to new physics producing spectral distortions with a characteristic momentum-transfer dependence. This is the case, for instance, of neutrino electromagnetic properties or light-mediator interactions; here we focus on the former. Using the recent XENONnT measurement in electron-recoil data, we first extract the normalizations of the $pp$ solar-neutrino component, validating our analysis against the official XENONnT one. 
Then, we provide the first determination of sin$^2\theta_W$ at $\sim 0.1$ MeV, and we derive bounds on the following electromagnetic properties: neutrino charge radii, magnetic moments, electric charges, and the sterile dipole portal.
Moreover, since the $pp$ flux inferred by XENONnT lies about $1.9\sigma$ above the expectation from standard solar models (SSMs), we also investigate the impact of a possible $^{14}$C background, whose spectrum closely mimics the solar-neutrino signal, on all of these implications. 
Our results demonstrate that the first observation of $pp$ solar neutrinos in a DM detector provides a powerful new laboratory for precision neutrino physics.

Our paper is organized as follows. In Sec.~\ref{sec:theory} we provide the relevant expression for the \eves~cross section both in the Standard Model and in presence of electromagnetic properties. We discuss in Sec.~\ref{sec:analysis} the procedure followed for the statistical analysis. We present in Sec.~\ref{sec:results} our results and in Sec.~\ref{sec:conclusions} we draw our conclusions.

%%%%%%%%%%%%%%%%%%%%%%%%%%%%%%%%%%%%%%%%%%%%%%%%%%%%%%%%%%%%%%%%
\section{Theoretical framework}
\label{sec:theory}
%%%%%%%%%%%%%%%%%%%%%%%%%%%%%%%%%%%%%%%%%%%%%%%%%%%%%%%%%%%%%%%%%%%%

In this section, we describe the \eves~cross sections relevant for the analysis. We first review the Standard Model (SM) prediction (Sec.~\ref{subsec:SMeves}) and then discuss how it is modified in the presence of beyond Standard Model (BSM) neutrino electromagnetic properties. 
In particular, we consider the neutrino charge radii (Sec.~\ref{subsec:chradii}), millicharges (Sec.~\ref{subsec:millich}), effects of neutrino magnetic moments (Sec.~\ref{subsec:magmom}) and sterile dipoles (Sec.~\ref{subsec:DP}). For more details and thorough reviews on the neutrino electromagnetic properties we refer the reader to Refs.~\cite{Giunti:2014ixa,Kouzakov:2017hbc,Giunti:2024gec}.

\subsection{E$\nu$ES in the Standard Model}
\label{subsec:SMeves}

In the SM, the differential cross section for \eves~of a neutrino $\nu_\ell$ with the bound electrons of an atom $\mathcal{A}$, as a function of the electron recoil energy $T_e$, is given by
\begin{align}
\label{eq:xsec_EvES_SM}
\begin{split}
\left. \frac{\d \sigma_{\nu_{\ell} \mathcal{A}}}{\d T_e}\right|^\mathrm{SM}= & Z^\mathcal{A}_\mathrm{eff}(T_e) \frac{G_F^2m_e}{2\pi}\left[(g^\ell_V + g^\ell_A)^2 +(g^\ell_V - g^\ell_A)^2\left(1-\frac{T_e}{E_\nu}\right)^2 -((g^\ell_V)^2-(g^\ell_A)^2)\frac{m_e T_e}{E_\nu^2}\right] \, .
\end{split}
\end{align}
Here, $G_F$ is the Fermi constant, while $E_\nu$ and $m_e$ denote the neutrino energy and the electron mass, respectively. The vector and axial-vector couplings depend on the flavor $\ell$ of the incoming neutrino and their tree-level values are given by
\begin{equation}
\label{eq:couplings}
g^\ell_V = 2\sin^2\theta_W-\frac{1}{2}+\delta_{\ell e},
\qquad
g^\ell_A = -\frac{1}{2}+\delta_{\ell e},
\end{equation}
where we take $\sin^2\theta_W = 0.23873$ for the weak mixing angle in the low-energy regime~\cite{ParticleDataGroup:2026mpi}. The term $\delta_{\ell e}$ accounts for the additional charged-current contribution that is present only for electron neutrinos. Thus, \eves~of $\nu_e$ receives contributions from both charged- and neutral-current interactions, whereas for $\nu_\mu$ and $\nu_\tau$ only the neutral-current contribution is present.

The target electrons are bound in atoms rather than being free. The atomic binding therefore modifies the scattering rate, particularly at low recoil energies. We account for these effects through $Z^\mathcal{A}_\mathrm{eff}(T_e)$, which is the effective number of electrons that can be ionized for a given deposited energy $T_e$.
At sufficiently high recoil energies, the atomic electrons can indeed be approximately treated as free particles, and the free-electron approximation (FEA) provides an adequate description of the scattering process\footnote{At sub-keV energies, atomic binding and many-electron effects can cause the FEA to overestimate the rate; see, e.g. the multiconfiguration relativistic random-phase approximation treatment in Ref.~\cite{Chen:2014ypv} and the implications for Xe in Ref.~\cite{Amaral:2026gml}.}.
Given the current electron recoil threshold reported by XENONnT, we therefore account for atomic binding and ionization effects through the effective number of electrons \(Z^\mathcal A_{\rm eff}(T_e)\), adopting the Hartree-Fock calculations of Ref.~\cite{Chen:2016eab}.

%%%%%%%%%%%%%%%%%%%%%%%%%%%%%%%%%%%%%%%%%%%%%%%%%%%%%%%%%%%%%%%
\subsection{Neutrino charge radii}
\label{subsec:chradii}

The neutrino charge radius (NCR) is the only electromagnetic neutrino parameter that is non-zero already within the SM. Even though neutrinos
are electrically neutral, the flavor-diagonal effective\footnote{Note that actually experiments are only sensitive to an effective charge radius $\langle r_\nu^2\rangle_{\rm eff} = \langle r_\nu^2\rangle - 6a_\nu$, which is a combination of charge radius $\langle r_\nu^2\rangle$ and anapole moment $a_\nu$.} charge radii are generated at one loop, giving~\cite{Bernabeu:2002pd,Bernabeu:2000hf,Bernabeu:2002nw}
\begin{equation}
\langle r^2_{\nu_{\ell\ell}} \rangle_\text{SM} =
-\frac{G_F}{2\sqrt{2}\pi^2}
\left[3 - 2\ln\frac{m_\ell^2}{M_W^2}\right] \, ,
\label{eq:CR_SM}
\end{equation}
where $m_\ell$ is the mass of the charged lepton of flavor
$\ell = e,\mu,\tau$ and $M_W$ is the $W$-boson mass. Numerically, one has
\begin{equation}
\left(\langle r^2_{\nu_{ee}}\rangle,\ \langle r^2_{\nu_{\mu\mu}}\rangle,\
\langle r^2_{\nu_{\tau\tau}}\rangle\right)_\text{SM} =
\left(-0.83,\ -0.48,\ -0.30\right)\times10^{-32}~\text{cm}^2 \, .
\label{eq:CR_SM_values}
\end{equation}
These flavor-diagonal charge radii are the only ones present in the SM; beyond the SM, neutrino mixing~\cite{Kouzakov:2017hbc} and other new-physics contributions~\cite{Cadeddu:2018dux} can additionally generate transition charge radii $\langle r^2_{\nu_{\ell\ell'}}\rangle$ with $\ell \neq \ell'$, which induce flavor transitions in the scattering process. In the following we consider the general case, allowing for both diagonal and transition charge radii.
 
The NCR contribution to the \eves~cross section is obtained through the substitution 
\begin{equation}
\label{eq:CR}
g_V^\ell \to g_V^\ell+Q_{\ell\ell}^{\rm NCR}, ~~\mathrm{with}~~
Q_{\ell\ell}^{\rm NCR}
=
\frac{\sqrt{2}\pi\alpha_{\rm EM}}
{3 G_F}
\langle r^2_{\nu_{\ell\ell}}\rangle\, ,
\end{equation}
where $\alpha_\mathrm{EM}$ is the fine-structure constant. The diagonal charge radii interfere coherently with the SM weak interaction, since both conserve flavor and helicity in the ultrarelativistic limit, whereas transition charge radii contribute incoherently. As a result, \eves~is sensitive to the diagonal charge radii $\langle r^2_{\nu_{\ell\ell}}\rangle$ including their sign, but only to the absolute values of the transition charge radii $\langle r^2_{\nu_{\ell\ell'}}\rangle$. In the latter case, the flavor-changing NCR cross section adds to the SM one and reads
\begin{equation}
\left. \frac{\d \sigma_{\nu_\ell \mathcal{A}}}{\d T_e}\right|^\mathrm{NCR}_{\ell\ell'}
=
\dfrac{ \pi \alpha^2_\mathrm{EM} m_{e}}{9}
\left[1 + \left( 1 - \dfrac{T_e}{E_\nu}\right)^2 - \dfrac{m_e T_e}{E_\nu^2} \right]
Z_\text{eff}^\mathcal{A}(T_e)
\left|\langle r^2_{\nu_{\ell\ell'}}\rangle\right|^2\, .
\label{eq:xsec_EvES_CR_FC}
\end{equation}
%

%%%%%%%%%%%%%%%%%%%%%%%%%%%%%%%%%%%%%%%%%%%%%%%%%%%%%%%%%%%%%%%%
\subsection{Neutrino electric charges}
\label{subsec:millich}

Another electromagnetic property of neutrinos that can be probed through \eves~(and also nuclear scattering) at direct detection experiments is a tiny electric charge (EC), commonly referred to as a neutrino \emph{millicharge}. 
In the SM, neutrinos are electrically neutral, but nonvanishing electric charges can arise in BSM extensions, including gauge models with right-handed neutrinos~\cite{Babu:1989tq}. More generally, the electromagnetic charge matrix can contain both flavor-diagonal charges, $q_{\ell \ell}$, and flavor-transition charges, $q_{\ell \ell'}$ with $\ell\neq\ell'$.

For flavor-diagonal millicharges, the additional contribution to \eves~can be incorporated into the SM cross section in Eq.~(\ref{eq:xsec_EvES_SM}) through the replacement
\begin{equation}
\label{eq:EC}
g_V^\ell \to g_V^\ell-Q_{\ell\ell}^{\rm EC}, ~~\mathrm{with}~~
Q_{\ell\ell}^{\rm EC}
=
\frac{\sqrt{2}\pi\alpha_{\rm EM}}
{G_F m_e T_e}
q_{\nu_{\ell\ell}}\,.
\end{equation}
For ultrarelativistic neutrinos, the electromagnetic charge interaction preserves helicity. Consequently, and in full analogy to the NCR case (\ref{subsec:chradii}), a flavor-diagonal millicharge preserves both helicity and flavor and therefore interferes coherently with the SM amplitude. This interference makes the corresponding scattering rate sensitive not only to the magnitude but also to the sign of the diagonal neutrino electric charge. In contrast, transition (i.e., flavor-changing) electric millicharges do not interfere with the flavor-conserving SM term. Their contribution enters only through the modulus squared of the transition charge
\begin{equation}
\left. \frac{\d \sigma_{\nu_\ell \mathcal{A}}}{\d T_e}\right|^\mathrm{EC}_{\ell\ell'}
=
\dfrac{ \pi \alpha^2_\mathrm{EM} }{m_{e} T_e^2}
\left[1 + \left( 1 - \dfrac{T_e}{E_\nu}\right)^2 - \dfrac{m_e T_e}{E_\nu^2} \right]
Z_\text{eff}^\mathcal{A}(T_e)
\left|q_{\nu_{\ell\ell'}} \right|^2\, .
\label{eq:xsec_EvES_millich_FC}
\end{equation}
Through the \eves~channel, the three flavor-diagonal charges, $q_{\nu_{ee}}$, $q_{\nu_{\mu\mu}}$, and $q_{\nu_{\tau\tau}}$, can in principle be constrained independently, while the three transition charges, $q_{\nu_{e\mu}}$, $q_{\nu_{e\tau}}$, and $q_{\nu_{\mu\tau}}$, are probed through their absolute values.

%%%%%%%%%%%%%%%%%%%%%%%%%%%%%%%%%%%%%%%%%%%%%%%%%%%%%%%%%%%%%%%%
\subsection{Neutrino magnetic moments}
\label{subsec:magmom}

Neutrino magnetic moments (MM) arise naturally in extensions of the SM that generate neutrino masses and have been constrained by a variety of laboratory, astrophysical, and cosmological observations; see, e.g., Ref.~\cite{Giunti:2024gec}.
Current limits, including the most stringent direct laboratory constraint by the GEMMA experiment~\cite{Beda:2012zz}, remain several orders of magnitude above the prediction in the minimal extension of the SM with right-handed neutrinos and Dirac masses~\cite{Fujikawa:1980yx,Pal:1981rm,Shrock:1982sc}. However, larger magnetic moments can arise in more general neutrino-mass models and other extensions of the SM~\cite{Babu:2020ivd}, motivating searches in current and future low-energy neutrino experiments~\cite{Giunti:2014ixa}. 
It is worth stressing that the effective magnetic moment is not a universal, experiment-independent quantity, but depends on the neutrino source, its flavor composition and energy spectrum and even on the Dirac or Majorana nature of neutrinos; consequently, the effective magnetic moment probed with solar neutrinos generally differs from that relevant for reactor or accelerator neutrino experiments~\cite{Ternes:2025lqh,AristizabalSierra:2021fuc}, and comparisons between bounds should be handled with care. 

A nonzero neutrino MM gives rise to an additional contribution to \eves~through the exchange of a photon. The MM interaction couples left- and right-handed neutrinos and therefore flips the neutrino helicity. Consequently, the MM \eves~cross section must be added incoherently to the SM one as they do not interfere and is given by
\begin{equation}
\left. \frac{\d \sigma_{\nu_\ell \mathcal{A}}}{\d T_e}\right|^\mathrm{MM}
=
\dfrac{ \pi \alpha^2_\mathrm{EM} }{ m_{e}^2 }
\left( \dfrac{1}{T_e} - \dfrac{1}{E_\nu} \right)
Z_\text{eff}^\mathcal{A}(T_e)
\left| \dfrac{\mu_{\nu_{\ell}}}{\mu_{\text{B}}} \right|^2 \, ,
\label{eq:xsec_EvES_magmom}
\end{equation}
where $\mu_B$ is the Bohr magneton. The parameter $\mu_{\nu_{\ell}}$ is the effective magnetic moment. It encodes the fundamental MM parameters and entries of the lepton mixing matrix~\cite{Canas:2015yoa,Miranda:2019wdy, AristizabalSierra:2021fuc,Ternes:2025lqh}.  Moreover, in fact it receives equal contributions from both the magnetic and electric dipole moments.
%

%%%%%%%%%%%%%%%%%%%%%%%%%%%%%%%%%%%%%%%%%%%%%%%%%%%%%%%%%%%%%%%%
\subsection{Sterile dipole portal}
\label{subsec:DP}

Sterile neutrinos are a well-motivated class of new-physics states, appearing naturally in a range of theoretical frameworks built to explain neutrino mass and mixing (see, e.g.,~\cite{Abdullahi:2022jlv}).
Here we focus on a simple such extension: the SM is extended by a single sterile state $\nu_4$ with mass $m_4$. When an active-sterile transition magnetic moment is present, it opens a new channel through which an active neutrino can convert into the sterile state by exchanging a photon with a target, $\nu_\ell + \mathcal{A} \to \nu_4 + \mathcal{A}$, a process known in the literature as the \textit{sterile dipole portal}~\cite{McKeen:2010rx,Magill:2018jla}. It is described by the effective Lagrangian~\cite{Gninenko:1998nn,Grimus:2000tq}
\begin{equation}
\mathcal{L}_\mathrm{DP}
=
\bar{\nu}_4(i\slashed{\partial}-m_4)\nu_4
+
\frac{\sqrt{\pi\alpha_\mathrm{EM}}}{2m_e}
\left|\frac{\mu_{\nu_\ell}^\mathrm{eff}}{\mu_{\mathrm{B}}}\right|
\bar{\nu}_4\sigma_{\mu\nu}\nu_\ell F^{\mu\nu}
+\,\mathrm{h.c.},
\end{equation}
with $F^{\mu\nu}$ the electromagnetic field-strength tensor and
$\sigma_{\mu\nu}=i[\gamma_\mu,\gamma_\nu]/2$. 
Since the dipole operator converts $\nu_\ell$ into a different final-state fermion, $\nu_4$, its amplitude does not interfere with the SM one at leading order, and its contribution adds incoherently to the SM rate\footnote{We neglect active-sterile mixing effects, which could induce
residual interference~\cite{Grimus:1997aa,Miranda:2021kre}.}.
The rate is the same whether $\nu_4$ is a Dirac or a Majorana state, since the active neutrino is purely left-handed; the two cases differ only in the subsequent $\nu_4\to\nu\gamma$ decay, which we assume to take place far outside the detector. Under this approximation, the differential \eves~cross section reads
\begin{equation}
\label{eq:xsec_EvES_dipolesterile}
\begin{aligned}
\left.\frac{\d\sigma_{\nu_\ell\mathcal{A}}}{\d T_e}\right|^\mathrm{DP}
={}&
\frac{\pi\alpha_\mathrm{EM}^2}{m_e^2}\,
Z_\mathrm{eff}^{\mathcal{A}}(T_e)
\left|\frac{\mu_{\nu_\ell}^\mathrm{eff}}{\mu_{\mathrm{B}}}\right|^2 \\[4pt]
&\times
\left[\frac{1}{T_e} -\frac{1}{E_\nu}
-\frac{m_4^2}{2E_\nu T_em_e}
\left(1-\frac{T_e}{2E_\nu} +\frac{m_e}{2E_\nu}\right)
+\frac{m_4^4(T_e-m_e)}{8 E_\nu^2 T_e^2 m_e^2}
\right] \, ,
\end{aligned}
\end{equation}
which reduces to the ordinary magnetic-moment term,
Eq.~\eqref{eq:xsec_EvES_magmom}, when $m_4\to0$. 
For $m_4\neq0$, upscattering requires $E_\nu \ge m_4 + m_4^2/(2m_e)$, and the $m_4$-dependent terms
suppress the rate, most strongly at low recoil energies.

%%%%%%%%%%%%%%%%%%%%%%%%%%%%%%%%%%%%%%%%%%%%%%%%%%%%%%%%%%%%%%%%%%%%
\section{Statistical analysis}
\label{sec:analysis}
%%%%%%%%%%%%%%%%%%%%%%%%%%%%%%%%%%%%%%%%%%%%%%%%%%%%%%%%%

In this section, we describe how we compute the expected electron-recoil spectrum induced by solar neutrinos in XENONnT, and the statistical procedure used to confront it with the data~\cite{XENON:2026rbz}. With an electron-recoil threshold of $\mathcal{O}(1)$~keV, the \eves~signal is dominated by the $pp$ and $^7$Be components of the solar flux, which together account for nearly all of the expected solar-neutrino
rate in the region of interest (ROI).
For each solar-neutrino component $k = pp,\, {}^7\mathrm{Be}$ and science run (SR $=$ SR0, SR1b), the differential event rate in true electron-recoil energy $T_e$ is
\begin{equation}
    \frac{\d R^k_\mathrm{SR}}{\d T_e}(\boldsymbol{\theta}) = \mathcal{E}_\mathrm{SR}  
    \int_{E_\nu^\text{min}(T_e)}^{E_\nu^\text{max}} \d E_\nu
    \sum_{\ell=e,\mu,\tau} \frac{\d\phi^k_\ell}{\d E_\nu}\,
    \frac{\d\sigma_{\nu_\ell}}{\d T_e} (\boldsymbol{\theta})\,,
    \label{eq:diff_rate}
\end{equation}
where $\d\sigma_{\nu_\ell}/\d T_e$ is the \eves\ cross section, which depends on the set $\boldsymbol{\theta}$ of parameters explored in each analysis (SM or one of the extensions of Sec.~\ref{sec:theory}), $\mathcal{E}_\mathrm{SR}$ is the exposure of the corresponding SR, $E_\nu^\text{max}$ is the endpoint of the flux, and $E_\nu^\text{min}(T_e) = \frac{1}{2}\big(T_e + \sqrt{T_e^2 + 2m_eT_e} \big)$ is the minimum neutrino energy\footnote{For the sterile dipole we have $E_\nu^\text{min, SD}(T_e) = E_\nu^\text{min}\times\left(1 + \frac{m_4^2}{2m_eT_e}\right)$.} required to produce a recoil $T_e$. 
The oscillated flux of each neutrino flavor reaching the detector is
\begin{equation}
    \frac{\d\phi^k_\ell}{\d E_\nu} =
    \frac{\d\phi^{k,0}_e}{\d E_\nu}\, P^k_{e\ell}(E_\nu)\,,
    \label{eq:osc_flux}
\end{equation}
where $\d\phi^{k,0}_e/\d E_\nu$ is the unoscillated $\nu_e$ flux, taken from the MB22m standard solar model~\cite{Magg:2022rxb,herrera_2023_10822316}, and $P^k_{e\ell}(E_\nu)$ is the $\nu_e \to \nu_\ell$ oscillation probability, averaged over the production region of component $k$ in the Sun. 
We fix the neutrino oscillation parameters to $\sin^2 \theta_{12}=0.309$, $\sin^2 \theta_{13}=0.022$, $\sin^2\theta_{23}=0.5$, $\Delta m_{21}^2=7.5 \times 10^{-5}~{\rm eV^2}$ and $\delta=0$ ~\cite{deSalas:2020pgw,JUNO:2025gmd} (see also Refs.~\cite{Esteban:2024eli,Capozzi:2025wyn}). Although maximal $\theta_{23}$ does not correspond to the best-fit value, this choice makes the $\nu_\mu$ and $\nu_\tau$ components of the solar flux at the detector equal, which simplifies the presentation of our constraints on flavor-dependent parameters. Let us stress, however, that varying these parameters within their allowed ranges has a negligible impact in our results.

The expected number of events in the $i$-th bin of reconstructed energy $T'_e$ is obtained by folding Eq.~\eqref{eq:diff_rate} with the detector energy resolution $A(T'_e, T_e)$~\cite{XENON2026announcement}, weighting by the published XENONnT efficiency $\mathcal{F}_\mathrm{SR}(T'_e)$~\cite{XENON:2026rbz}, and integrating over the bin,
\begin{equation}
    R^k_{i,\mathrm{SR}}(\boldsymbol{\theta}) = \int_{T'_{i,\text{min}}}^{T'_{i,\text{max}}} \d T'_e\,
    \mathcal{F}_\mathrm{SR}(T'_e) \int \d T_e\, A(T'_e, T_e)\,
    \frac{\d R^k_\mathrm{SR}}{\d T_e}(\boldsymbol{\theta})\,.
    \label{eq:binned_rate}
\end{equation}
For each SR, we divide the ROI, $T'_e \in [1, 140]$~keV, into 139 evenly spaced bins of 1~keV width.
The total number of events predicted in each bin $i$ of a given SR is given by
\begin{equation}
    N_{i,\mathrm{SR}}(\boldsymbol{\theta}) =
    \sum_{j= pp, {}^7\textrm{Be}} R_{i,\mathrm{SR}}^{j}(\boldsymbol{\theta})\,(1+\alpha_j)
    + \sum_b B^b_{i,\mathrm{SR}}\,(1+\beta^b_\mathrm{SR})\,,
    \label{eq:total_in_bin}
\end{equation}
where $B^b_{i,\mathrm{SR}}$ is the rate of background component $b$ in bin $i$. The spectrum of each background component, together with its normalization uncertainty, is provided by the Collaboration for each SR~\cite{XENON:2026rbz}.
The nuisance parameters $\alpha_{pp}$ and $\alpha_{{}^{7}\mathrm{Be}}$ account for the uncertainty on the solar flux predictions, $\sigma_{pp} = 0.59\%$ and $\sigma_{{}^{7}\mathrm{Be}} = 7.64\%$~\cite{Magg:2022rxb,herrera_2023_10822316}, while $\beta^k_\mathrm{SR}$ rescale the background normalizations, with uncertainties $\sigma_{\beta^b_\mathrm{SR}}$ provided in~\cite{XENON:2026rbz}. The background normalizations are varied independently in each SR, except for the $^{124}$Xe double electron capture and $^{136}$Xe double-beta decay components. These are intrinsic to the xenon target and are therefore treated as nuisance parameters shared between the two SRs.

For a given physics scenario, characterized by the SM or BSM parameter(s), $\boldsymbol{\theta}$, our statistical analysis is performed by comparing our predictions with the measured counts $D_{i,\mathrm{SR}}$ through the Poissonian statistic
\begin{equation}
    \chi^2_\mathrm{SR}(\boldsymbol{\theta}) = 2 \sum_i \left[ N_{i,\mathrm{SR}}(\boldsymbol{\theta}) - D_{i,\mathrm{SR}}
    + D_{i,\mathrm{SR}} \ln\!\left(\frac{D_{i,\mathrm{SR}}}{N_{i,\mathrm{SR}}(\boldsymbol{\theta})}\right)
    \right]\,.
    \label{eq:chi2_SR}
\end{equation}
Combining both SRs with the nuisance-parameter penalties gives the following expression for the profiled statistic
\begin{equation}
    \chi^2(\boldsymbol{\theta}) =
    \min_{\alpha_{pp},\,\alpha_{{}^{7}\mathrm{Be}},\,\vec\beta}\,
    \left\{\chi^2_{\rm SR0}(\boldsymbol{\theta}) + \chi^2_{\rm SR1b} (\boldsymbol{\theta}) 
    + \left(\frac{\alpha_{pp}}{\sigma_{pp}}\right)^2
    + \left(\frac{\alpha_{{}^{7}\mathrm{Be}}}{\sigma_{{}^{7}\mathrm{Be}}}\right)^2
    + \sum_{\mathrm{SR}} \sum_b
      \left(\frac{\beta^b_\mathrm{SR}}{\sigma_{\beta^b_\mathrm{SR}}}\right)^2\right\}\, ,
    \label{eq:chi2_analysis}
\end{equation}
where the penalties of the shared nuisance parameters are counted only once. This expression, however, does not yet include the systematic uncertainties affecting the solar-neutrino signal, to which we now turn.

\textbf{Systematic uncertainties.} Beyond the flux and background normalizations, the XENON Collaboration reports a set of systematic uncertainties affecting the solar-neutrino signal (see Tab.~III of Ref.~\cite{XENON:2026rbz}). Adding the lower and upper contributions in quadrature gives $\sigma_\mathrm{sys}^{-} = 14.35\%$ and $\sigma_\mathrm{sys}^{+} = 14.04\%$. Given the small asymmetry, we adopt their average, $\sigma_\mathrm{sys} = 14.19\%$.

We implement this systematic uncertainty as a common rescaling of the total solar-neutrino yield that preserves its spectral shape.  For later convenience, we define the unshifted total solar-neutrino yield as 
\begin{equation}
    N_\odot
    \left(
        \boldsymbol{\theta},
        \alpha_{pp},
        \alpha_{{}^{7}{\rm Be}}
    \right)
    =
    \sum_{{\rm SR}}
    \sum_i
    \left[
        R_{i,{\rm SR}}^{pp}(\boldsymbol{\theta})(1+\alpha_{pp})
        +
        R_{i,{\rm SR}}^{{}^{7}{\rm Be}}(\boldsymbol{\theta})
        (1+\alpha_{{}^{7}{\rm Be}})
    \right]\,.
    \label{eq:totsolar_yield}
\end{equation}
To account for systematic effects in the predicted solar \eves~signal, we allow for a shift $N_\odot \to N_\odot + \Delta N$ in the expected event rate, corresponding to an overall rescaling of the solar contribution by
\begin{equation}
    f_{\Delta N}
    =
    \frac{N_\odot+\Delta N}{N_\odot}
    =
    1+\frac{\Delta N}{N_\odot}\,.
\end{equation}
The expected number of events in each SR and energy bin therefore becomes
\begin{equation}
    N_{i,{\rm SR}}(\Delta N)
    =
    f_{\Delta N}
    \left[
        R_{i,{\rm SR}}^{pp}(\boldsymbol{\theta})(1+\alpha_{pp})
        +
        R_{i,{\rm SR}}^{{}^{7}{\rm Be}}(\boldsymbol{\theta})
        (1+\alpha_{{}^{7}{\rm Be}})
    \right]
    +
    \sum_k B_{i,{\rm SR}}^k(1+\beta^k_\mathrm{SR})\,.
\end{equation}
For each fixed $(\boldsymbol{\theta}, \Delta N)$, we define $\chi^2_\mathrm{prof}(\boldsymbol{\theta}, \Delta N)$ as Eq.~\eqref{eq:chi2_analysis} with $N_{i,\mathrm{SR}}(\boldsymbol{\theta})$ replaced by $N_{i,\mathrm{SR}}(\boldsymbol{\theta}, \Delta N)$, i.e. profiling over $\alpha_{pp}$, $\alpha_{{}^{7}{\rm Be}}$ and $\vec\beta$. We then perform a post-fit convolution by integrating the profiled likelihood over $\Delta N$, weighted by a Gaussian prior

\begin{equation}
    \mathcal{L}_{\rm conv}(\boldsymbol{\theta})
    =
    \frac{1}{\sqrt{2\pi}\sigma_N}
    \int \d(\Delta N)\,
    \exp\left[
        -\frac{(\Delta N)^2}{2\sigma_N^2}
    \right]
    \exp\left[
        -\frac{1}{2}
        \chi^2_{\rm prof}(\boldsymbol{\theta},\Delta N)
    \right]\,.
\end{equation}
The width is fixed to $\sigma_N = \sigma_\mathrm{sys}\,N_\odot^\mathrm{BF}$, where $N_\odot^{\rm BF} \equiv N_\odot
    \left(
        \boldsymbol{\theta}^{\rm BF},
        \alpha_{pp}^{\rm BF},
        \alpha_{{}^{7}{\rm Be}}^{\rm BF}
    \right)$    
is given by Eq.~(\ref{eq:totsolar_yield}) evaluated at the best-fit values obtained from Eq.~(\ref{eq:chi2_analysis}).
The statistic used to derive our results, including the additional systematics, is then
\begin{equation}
    \chi^2_{\rm conv}(\boldsymbol{\theta})
    =
    -2\ln\mathcal{L}_{\rm conv}(\boldsymbol{\theta})\, .
\end{equation}

\textbf{The $^{14}$C background.} Finally, we note that the $pp$ flux inferred by the XENON Collaboration in their own analysis~\cite{XENON:2026rbz} is larger than, and about $1.9\sigma$ away from, the previous measurement by Borexino~\cite{BOREXINO:2018ohr} and about $2\sigma$ from the theoretical expectation assuming SSMs~\cite{Magg:2022rxb}. Motivated by this mild tension, we investigate the role of the $^{14}$C background. Its $\beta$-spectrum closely resembles that of the solar-neutrino signal in the ROI, so an unaccounted $^{14}$C component would be largely absorbed into a larger $pp$ flux. The most plausible carrier of residual $^{14}$C in the xenon is CO$_2$, into which cosmogenic $^{14}$C, produced in the upper atmosphere, is predominantly incorporated. In an appendix of Ref.~\cite{XENON:2026rbz},
the XENON Collaboration includes $^{14}$C as an additional background, constraining the CO$_2$ concentration to $[0, 100]$~ppt. 
As explained by the Collaboration, this range is set by the quoted upper bound on CO$_2$ removal by the gaseous purification system, while outgassing from detector materials was found to be negligible. 
Here we go one step further and treat $^{14}$C as an independent background component in each SR: we keep the $[0, 100]$~ppt constraint
on the CO$_2$ concentration in SR0 but leave the $^{14}$C normalization free in SR1b. Our motivation is that, despite the additional distillation campaign and the exclusion of SR1a data, residual $\rm CO_2$ contamination could have entered the detector, preventing its concentration in SR1b from returning to the SR0 level. In both SRs, the $^{14}$C normalization enters without a Gaussian penalty in Eq.~\eqref{eq:chi2_analysis}. Its impact on our results is discussed in Sec.~\ref{sec:results}.

%%%%%%%%%%%%%%%%%%%%%%%%%%%%%%%%%%%%%%%%%%%%%%%%%%%%%%%%%
\section{Results}
\label{sec:results}
%%%%%%%%%%%%%%%%%%%%%%%%%%%%%%%%%%%%%%%%%%%%%%%%%%%%%%%%%

We now present the results of our analyses. We first examine the implications for SM physics: the normalization of the $pp$ solar-neutrino flux, and the weak mixing angle $\sin^2\theta_W$ at low momentum transfer.
We then turn to the sensitivity to the electromagnetic properties of neutrinos: the charge radius, a possible
millicharge, the magnetic moment and, finally, the sterile dipole portal. Throughout, we show results both with and without the $^{14}$C background discussed in Sec.~\ref{sec:analysis}.

%%%%%%%%%%%%%%%%%%%%%%%%%%%%%%%%%%%%%%%%%%%%%%%%%%%%%%%%%
\subsection{SM physics: $pp$ and $^7$Be flux normalization and weak mixing angle}
\label{sec:res-SM}

\begin{figure}[t!]
\centering
\includegraphics[width=0.42\textwidth]{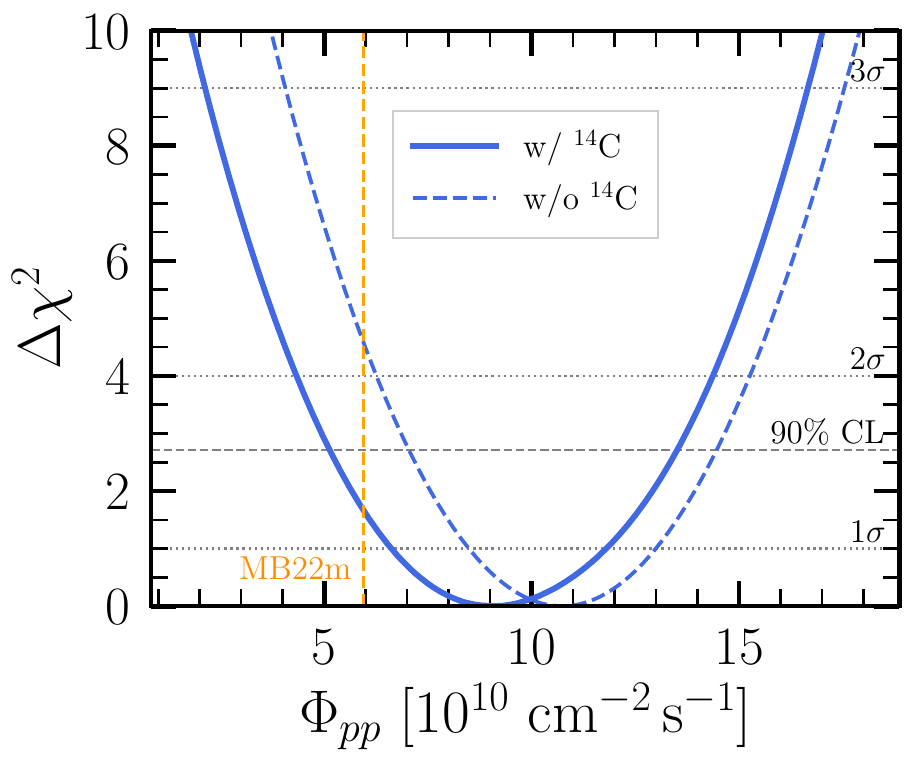}
\includegraphics[width=0.5\textwidth]{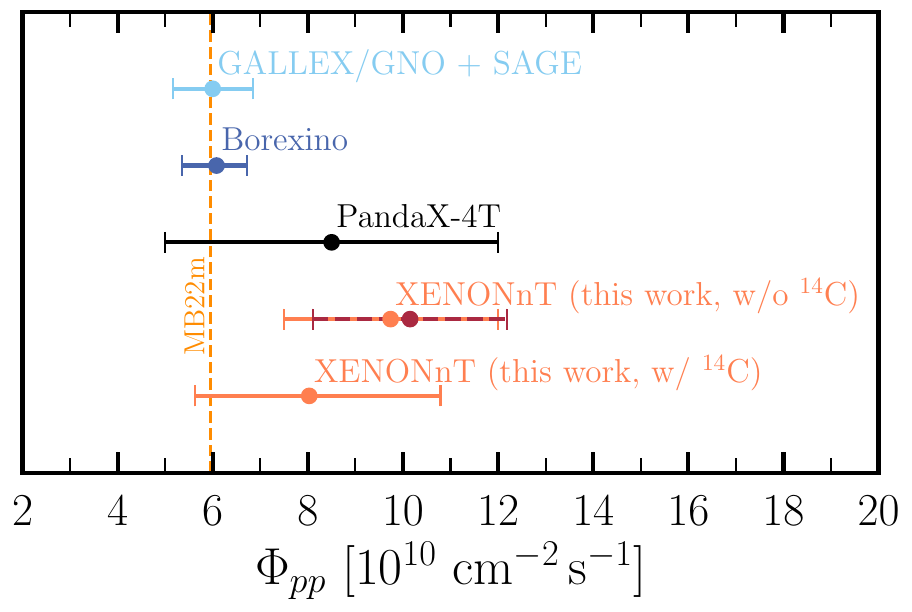}

\caption{Left: Profiled $\Delta\chi^2$ for the $pp$ solar-neutrino flux, with (solid) and without (dashed) the $^{14}$C background. The orange dashed vertical line marks the MB22m prediction. Right: Determinations of the solar $pp$ neutrino flux, with $1\sigma$ error bars. The results of this work, obtained from the XENONnT data~\cite{XENON:2026rbz} without (upper) and with (lower) the $^{14}$C background, are compared with the published XENONnT result (dashed), PandaX-4T~\cite{PandaX:2026qdp}, Borexino~\cite{BOREXINO:2018ohr}, and the values inferred from the radiochemical gallium experiments (GALLEX/GNO~\cite{Kaether:2010ag} and SAGE~\cite{SAGE:2009eeu}). The dashed vertical line indicates the MB22m Standard Solar Model prediction.}
\label{fig:fluxes}
\end{figure}

We begin with a purely SM interpretation of our results and we first determine the solar-neutrino flux normalization. To this end, we fix all SM couplings to their standard (tree-level) values and treat the $pp$ solar flux as the parameter of interest, while constraining the complementary ${}^7$Be fluxes within the MB22m theoretical uncertainty~\cite{Magg:2022rxb,herrera_2023_10822316}.
We note that the two main flux components ($pp$ and ${}^7$Be) cannot both be left free simultaneously, for two reasons. First, in the ROI they produce electron-recoil spectra of very similar shape: the $^7$Be lines yield a nearly flat spectrum, and the low-energy $pp$ spectrum is also smooth. A simultaneous fit is therefore almost entirely degenerate, with the data constraining mainly the total solar-neutrino rate. Second, both fluxes originate from the $pp$ chain, so that a luminosity-weighted combination of the two is fixed by the measured solar luminosity: an increase of the $^7$Be flux must therefore be compensated by a decrease of the $pp$ flux. Constraining one flux with the solar-model prediction breaks the degeneracy while keeping the fit consistent with solar physics.
We obtain at $1\sigma$
\begin{align}
\label{eq:res_pp}
\Phi_{pp}^{\rm w/o\,{}^{14}C}
&=
\left(10.74^{+2.26}_{-2.25}\right)
\times 10^{10}~{\rm cm}^{-2}\,{\rm s}^{-1},
\nonumber\\
\Phi_{pp}^{\rm w/\,{}^{14}C}
&=
\left(9.03^{+2.76}_{-2.41}\right)
\times 10^{10}~{\rm cm}^{-2}\,{\rm s}^{-1}.
\end{align}
Without the $^{14}$C background, the determination lies above the corresponding MB22m solar prediction $\Phi_{pp}^{\rm MB22m} = 5.95\times10^{10}~{\rm cm}^{-2}\,{\rm s}^{-1}$, at the $2.1\sigma$ level.
When the hypothetical $^{14}$C background is included, part of the excess is absorbed by this component, and the agreement with the nominal values improves to $1.3\sigma$. In particular, including the $^{14}$C background lowers our obtained $pp$ flux by $1.7\times10^{10}$~cm$^{-2}$\,s$^{-1}$, comparable to the $1.5\times10^{10}$~cm$^{-2}$\,s$^{-1}$ shift found by XENONnT~\cite{XENON:2026rbz}. As shown in Appendix~\ref{app:flux_validation}, the small difference stems from the different solar-model inputs and fit setup, rather than from our treatment of the SR1b $^{14}$C normalization. 
The corresponding profiled $\Delta\chi^2$ curves are shown in the left panel of Fig.~\ref{fig:fluxes}, which shows that including the $^{14}$C background moves the minimum to a lower value, towards the MB22m predictions. 
Our result is also consistent with the independent reanalysis of
Ref.~\cite{Amaral:2026vkt}, which finds $\Phi_{pp} = (10.3\pm1.2)\times10^{10}$~cm$^{-2}$\,s$^{-1}$; the smaller uncertainty quoted there is consistent with the one we obtain when neglecting the signal systematic uncertainty $\sigma_{\rm sys}$
(see Appendix~\ref{app:flux_validation}).

In the right panel of Fig.~\ref{fig:fluxes} we compare our $pp$ flux determinations (in orange, solid) with previous measurements. Although the precision is still far from that of dedicated solar-neutrino experiments, these results illustrate that dark matter detectors are starting to measure low-energy solar-neutrino fluxes directly, establishing them as complementary tools for solar-neutrino studies.
In Appendix~\ref{app:flux_validation}, we detail the validation of our flux analysis against the nominal XENONnT one, and discuss the impact of our treatment of the $^{14}$C concentration, compared to the Collaboration's prescription.

\begin{figure}[t]
\centering
\includegraphics[width=0.58\textwidth]{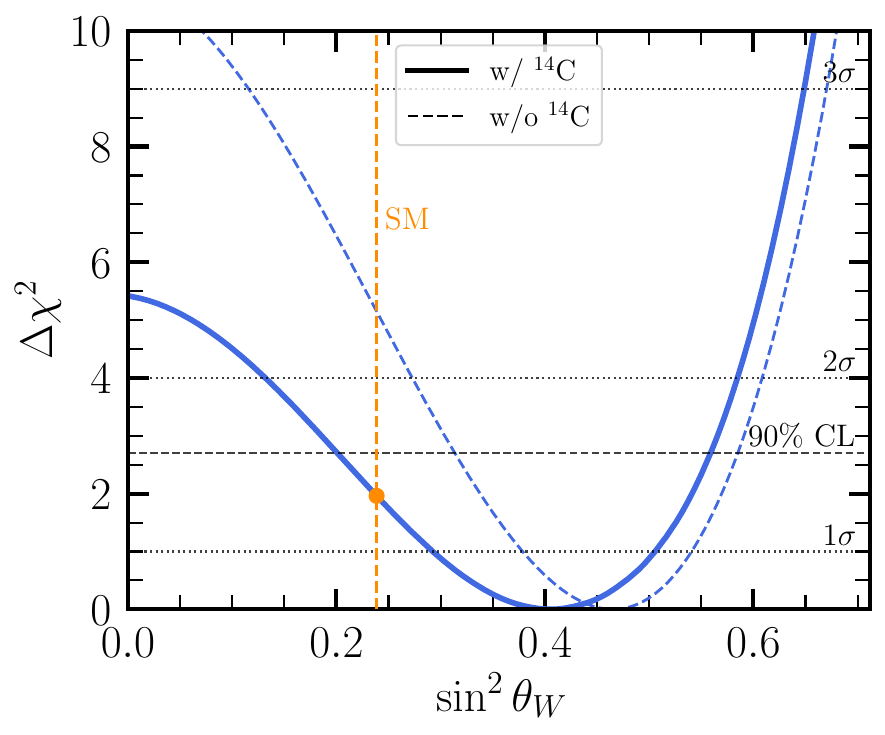}
\caption{Profiled $\Delta\chi^2$ for the weak mixing angle $\sin^2\theta_W$, with (blue, solid) and without (blue, dashed) the $^{14}$C background. The orange, dashed vertical line marks the SM prediction, as defined in the $\overline{\rm MS}$ renormalization scheme~\cite{Erler:2017knj,Erler:2004in}.}
\label{fig:sw2}
\end{figure}

To determine the weak mixing angle, we fix both $\Phi_{pp}$ and $\Phi_{{}^{7}{\rm Be}}$ to their theoretical predictions, and treat $\sin^2\theta_W$ as the parameter of interest, profiling over the remaining background and solar-flux nuisance parameters. This isolates the sensitivity of the electron-recoil spectrum to the neutral-current vector coupling, which enters the \eves~cross section via $g_V^\ell$ in Eq.~\eqref{eq:couplings}.
At these low recoil energies, the weak mixing angle is probed at a momentum transfer that ranges from $\sim0.03$~MeV at the $T_e = 1$~keV threshold to $\sim0.4$~MeV at the upper edge of the ROI. This lies below both the scale of atomic parity violation (APV) in Cesium ($\qtransfer \simeq 2.4$~MeV~\cite{ParticleDataGroup:2026mpi}), and that of solar and terrestrial \cevns~measurements ($\qtransfer \sim 10-100$ MeV) \cite{DeRomeri:2022twg,Majumdar:2022nby,AtzoriCorona:2022qrf,DeRomeri:2025csu,DeRomeri:2026dac,DeRomeri:2026prc,Maity:2024aji,AtzoriCorona:2025gyz,AtzoriCorona:2026wbu,Ahmed:2026knn}. 

The resulting profiles are shown in Fig.~\ref{fig:sw2}. At $1\sigma$ we report the best-fit values
\begin{align}
\sin^2\theta_W^{\rm w/o\,{}^{14}C}
&=
0.465^{+0.075}_{-0.087}\,,
\nonumber\\
\sin^2\theta_W^{\rm w/\,{}^{14}C}
&=
0.407^{+0.099}_{-0.115}\,.
\end{align}
The corresponding $90\%$ confidence level (CL) intervals are $[0.313,\,0.585]$ and $[0.200,\,0.559]$ for the analyses without and with the $^{14}$C contribution, respectively. Both best-fit values lie well above the SM prediction at low momentum transfer, $\sin^2\theta_W = 0.23873$~\cite{ParticleDataGroup:2026mpi}, defined in the $\overline{\rm MS}$ renormalization scheme~\cite{Erler:2004in,Erler:2017knj}. The SM value corresponds to $\Delta\chi^2 = 5.15$ ($\simeq2.3\sigma$) without the $^{14}$C background and to $\Delta\chi^2 = 1.96$ ($\simeq1.4\sigma$) when it is included. This value depends only mildly on our treatment of the $^{14}$C background in SR1b, with unchanged best fit (see Appendix~\ref{app:flux_validation}).

This behavior is expected: with the solar fluxes constrained to their SSM values, the same excess of events that drives the large $pp$ flux in Eq.~\eqref{eq:res_pp} must now be absorbed by an enhanced neutral-current coupling, pushing $\sin^2\theta_W$ well above its SM prediction at that energy scale.  The reduced tension once $^{14}$C is included suggests that the excess may, at least in part, originate from an imperfect modelling of the background at low recoil energies, rather than a genuine deviation from the SM; a more refined background treatment could bring the extracted $\sin^2\theta_W$ into better agreement with its SM expectation. 

We can compare these results with recent determinations using the \cevns~channel, which probe $\sin^2\theta_W$ at higher momentum transfer.
Solar $^8$B \cevns\ data yield a combined value of $0.20^{+0.05}_{-0.04}$~\cite{DeRomeri:2025csu}, while recent reactor and accelerator measurements give $\sin^2\theta_W = 0.235^{+0.025}_{-0.024}$ (COHERENT-Ge) and $0.247^{+0.052}_{-0.051}$ (CONUS+)~\cite{DeRomeri:2026dac} (see also~\cite{AtzoriCorona:2026wbu,CONUS:2026uhz,Ahmed:2026knn}). All of these are consistent with the SM running. Our $1\sigma$ uncertainty is comparable to that of the individual solar \cevns~determinations, but about three times larger than that of COHERENT-Ge, while our central value lies above all of them. 
Since the \cevns~determinations rely on nuclear recoils and are independent of both the $pp$ flux and the electron-recoil backgrounds, we regard this comparison as further indication that the high value of $\sin^2\theta_W$ reflects the normalization of the electron-recoil signal rather than a genuine low-energy running effect. Let us also note that earlier DM electron-recoil data, without a $pp$ detection, implied lower values at similar momentum transfer,  $\sim 0.27^{+0.11}_{-0.18}$ according to Ref.~\cite{AtzoriCorona:2026lqk}, while the global fit of low-energy
neutrino data in the same reference yields $0.233^{+0.007}_{-0.006}$, in agreement with the SM.

The same excess currently prevents a meaningful extraction of the solar electron-neutrino survival probability. Notice that even if we considered $P_{ee}=1$, the rate would increase by only $\sim50\%$ with respect to the standard expectation, $P_{ee}\simeq 0.55$, less than the~80\% excess implied by Eq.~\eqref{eq:res_pp}. 
A fit to $P_{ee}$ therefore prefers unphysical values, $P_{ee}>1$, and even with the $^{14}$C background included, the large statistical uncertainties leave $P_{ee}$ essentially unconstrained within the physical range $[0,1]$. A determination of $P_{ee}$ at $pp$ energies will require larger exposure and a better understanding of the low-energy backgrounds~\cite{DARWIN:2020bnc}.

%%%%%%%%%%%%%%%%%%%%%%%%%%%%%%%%%%%%%%%%%%%%%%%%%%%%%%%%%
\subsection{Neutrino charge radii}
\label{subsec:res_CR}

% ============================================================
% CHARGE RADIUS: 2D
% ===========================================================

\begin{figure}[t]
\centering

\includegraphics[width=\textwidth]{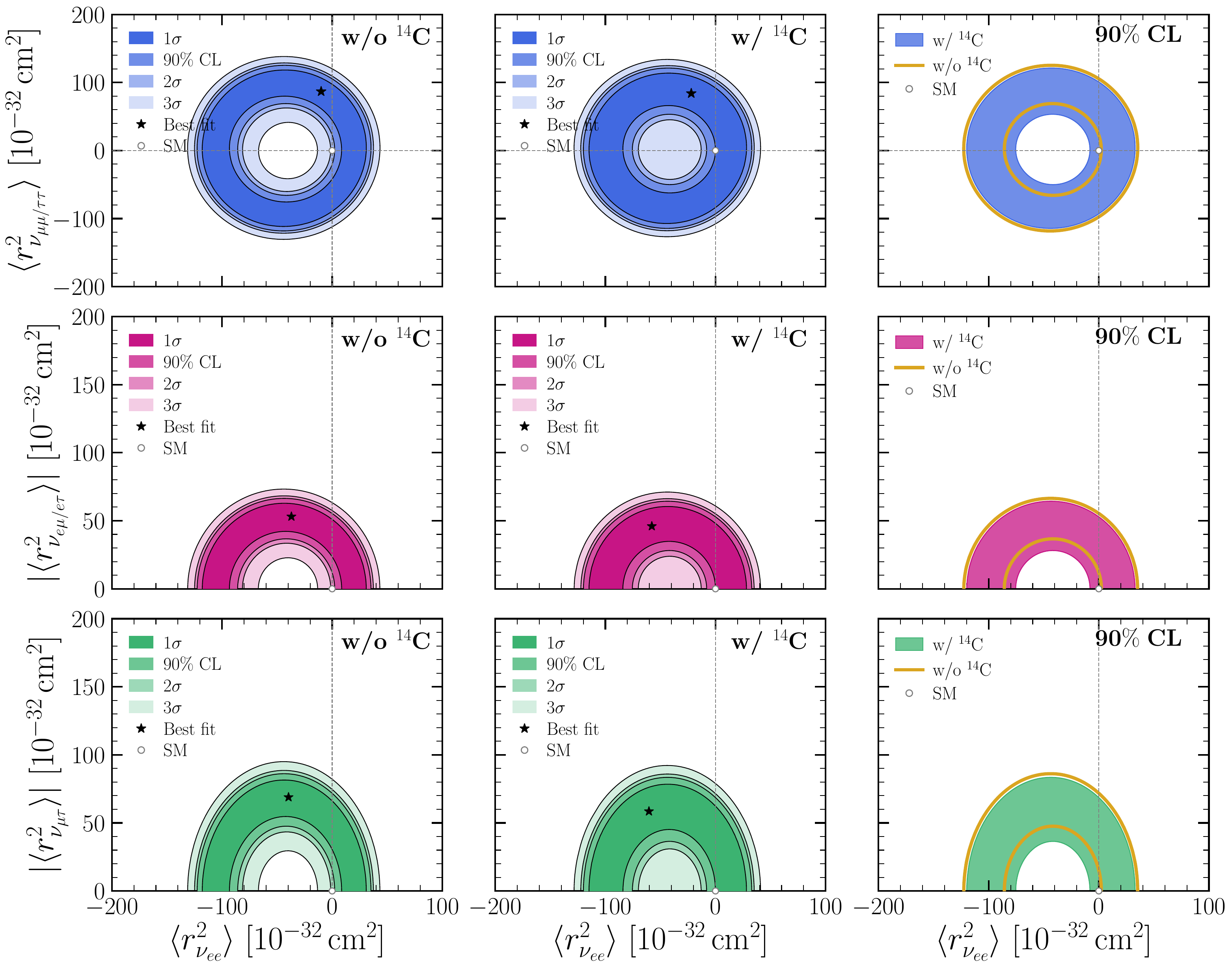}

\caption{Two-dimensional constraints on the neutrino charge radii in the $\langle r^2_{\nu_{ee}}\rangle-\langle r^2_{\nu_{\mu\mu/\tau\tau}}\rangle$ (top row), $\langle r^2_{\nu_{ee}}\rangle-|\langle r^2_{\nu_{e\mu/e\tau}}\rangle|$ (middle row), and $\langle r^2_{\nu_{ee}}\rangle-|\langle r^2_{\nu_{\mu\tau}}\rangle|$ (bottom row) planes, with the charge radii not shown profiled over. The
left and central columns show the results without and with the $^{14}$C background, respectively, at $1\sigma$, $90\%$, $2\sigma$ and $3\sigma$ CL;  stars mark the best-fit points, while the open circle and dashed lines the SM prediction. The right column compares the $90\%$ CL regions with (filled) and without (orange contour) the $^{14}$C background.}
\label{fig:charge_radius_2d}
\end{figure}

% ============================================================
% CHARGE RADIUS: PROFILED MULTIPARAMETER RESULTS
% ============================================================

\begin{figure}[t]
\centering
\includegraphics[width=0.49\textwidth]{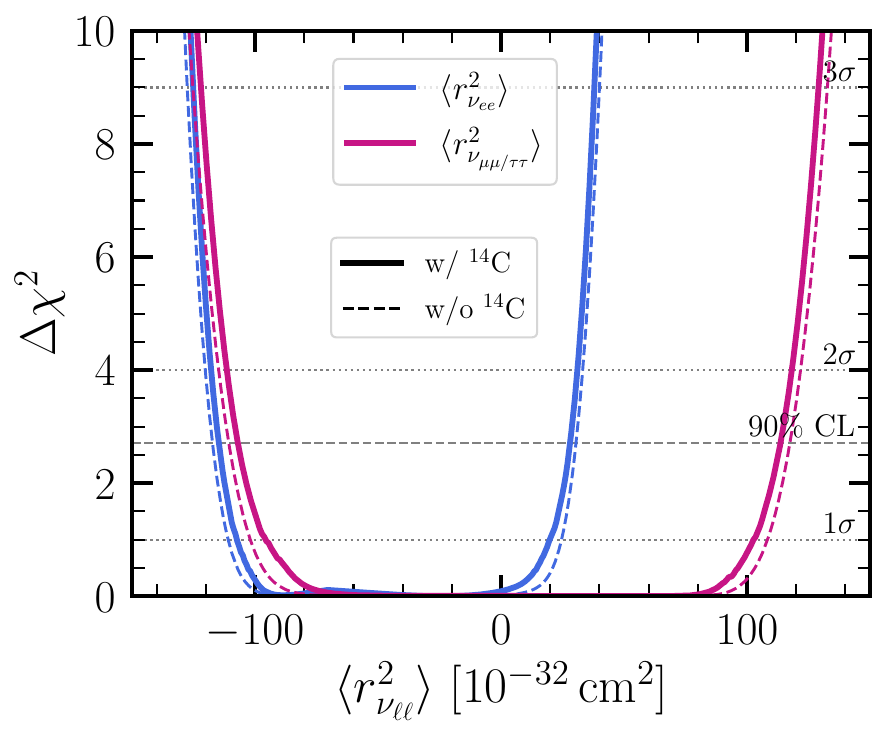}
\includegraphics[width=0.49\textwidth]{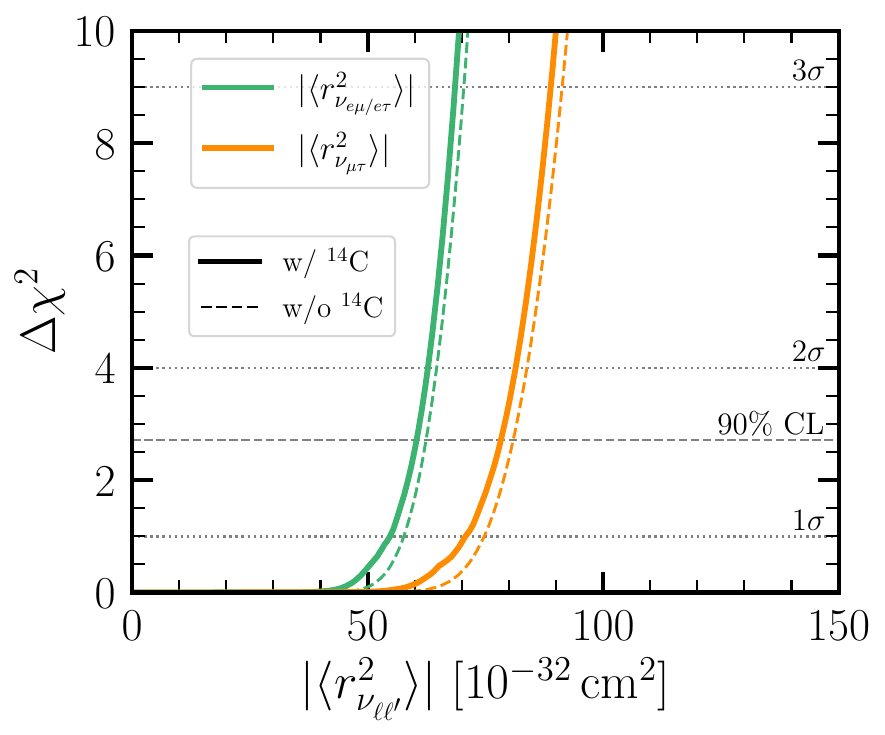}
\caption{Profiled $\Delta\chi^2$ for the neutrino charge radii, with all other charge radii profiled over. Left: flavor-diagonal charge radii.
Right: transition charge radii. Solid and dashed curves correspond to the analyses with and without the $^{14}$C background, respectively.}
\label{fig:charge_radius_profiled}
\end{figure}

We now turn to the electromagnetic properties of neutrinos, starting with the charge radii. 
Since we have chosen $\sin^2\theta_{23}=0.5$ and since $\nu_\mu$ and $\nu_\tau$ have identical \eves~cross sections, the data cannot distinguish between $\langle r^2_{\nu_{\mu\mu}}\rangle$ and $\langle r^2_{\nu_{\tau\tau}}\rangle$,
nor between $\langle r^2_{\nu_{e\mu}}\rangle$ and $\langle r^2_{\nu_{e\tau}}\rangle$. We therefore present results for
$\langle r^2_{\nu_{ee}}\rangle$, $\langle r^2_{\nu_{\mu\mu/\tau\tau}}\rangle$, $|\langle r^2_{\nu_{e\mu/e\tau}}\rangle|$ and $|\langle r^2_{\nu_{\mu\tau}}\rangle|$.

Figure~\ref{fig:charge_radius_2d} shows the allowed regions in three two-dimensional projections of this parameter space, with the remaining charge radii profiled over, while Fig.~\ref{fig:charge_radius_profiled} shows the corresponding one-dimensional profiles.
The left and central columns of Fig.~\ref{fig:charge_radius_2d} show the results without and with the $^{14}$C background, respectively, at $1\sigma$, $90\%$, $2\sigma$, and $3\sigma$; the right column compares the corresponding $90\%$ CL regions. 
At $1\sigma$ we obtain 
\begin{align}
\langle r^2_{\nu_{ee}}\rangle^{\rm w/o\,{}^{14}C}
&\in
[-110.94,\,24.59]\times10^{-32}\,{\rm cm}^2,
&
\langle r^2_{\nu_{ee}}\rangle^{\rm w/\,{}^{14}C}
&\in
[-107.40,\,19.80]\times10^{-32}\,{\rm cm}^2,
\nonumber\\
\langle r^2_{\nu_{\mu\mu/\tau\tau}}\rangle^{\rm w/o\,{}^{14}C}
&\in
[-101.96,\,108.32]\times10^{-32}\,{\rm cm}^2,
&
\langle r^2_{\nu_{\mu\mu/\tau\tau}}\rangle^{\rm w/\,{}^{14}C}
&\in
[-95.71,\,102.62]\times10^{-32}\,{\rm cm}^2,
\nonumber\\
|\langle r^2_{\nu_{e\mu/e\tau}}\rangle|^{\rm w/o\,{}^{14}C}
&<
57.69\times10^{-32}\,{\rm cm}^2,
&
|\langle r^2_{\nu_{e\mu/e\tau}}\rangle|^{\rm w/\,{}^{14}C}
&<
54.77\times10^{-32}\,{\rm cm}^2,
\nonumber\\
|\langle r^2_{\nu_{\mu\tau}}\rangle|^{\rm w/o\,{}^{14}C}
&<
74.77\times10^{-32}\,{\rm cm}^2,
&
|\langle r^2_{\nu_{\mu\tau}}\rangle|^{\rm w/\,{}^{14}C}
&<
70.75\times10^{-32}\,{\rm cm}^2.
\end{align}
The corresponding $90\%$ CL intervals are listed in Tab.~\ref{tab:em_limits}. We quote allowed intervals rather than best-fit values, since, as discussed below, the profiles are nearly flat over most of these intervals.

The shape of the allowed regions can be understood from the way the charge radii enter the cross section. As discussed in Sec.~\ref{subsec:chradii}, a diagonal charge radius shifts the vector coupling, $g_V^\ell \to g_V^\ell + Q^{\rm NCR}_{\ell\ell}$, independently of the recoil energy, so that the data are mainly sensitive to the overall rate. The diagonal contributions enter through $(g_V^\ell + Q^{\rm NCR}_{\ell\ell})^2$, while the transition charge radii contribute incoherently, through $|\langle r^2_{\nu_{\ell\ell'}}\rangle|^2$. Curves of constant rate are therefore ellipses centered at $Q^{\rm NCR}_{\ell\ell} = -g_V^\ell$ and vanishing transition charge radii, and a measurement of the rate selects the elliptical rings visible in Fig.~\ref{fig:charge_radius_2d}. Along each ring the rate, and hence the $\chi^2$, is approximately constant, so the position of the best-fit points is not physically meaningful. For $\nu_e$, the coupling $g_V^e = \tfrac12 + 2\sin^2\theta_W \simeq 0.98$ includes the charged-current contribution, and the center of the rings lies at $\langle r^2_{\nu_{ee}}\rangle \simeq -41\times10^{-32}$~cm$^2$, and the allowed interval extends more towards negative rather than positive values. For $\nu_{\mu,\tau}$, instead, $g_V^\mu$ is accidentally close to zero, so the rings, and hence the interval on $\langle r^2_{\nu_{\mu\mu/\tau\tau}}\rangle$, are nearly symmetric around the origin.

The interior of the rings corresponds to an \eves~rate below the SM one and is excluded by the observed excess at more than $3\sigma$ without the $^{14}$C background; the SM point lies within the $1\sigma$ region. The results are nonetheless robust against the
inclusion of $^{14}$C: as shown in the right column of
Fig.~\ref{fig:charge_radius_2d}. The outer boundaries of the $90\%$ CL regions are almost unchanged. The main effect is on the inner boundaries: since part of the excess is absorbed by the $^{14}$C background, lower rates become allowed, and the interior of the rings is no longer excluded at $3\sigma$.

All bounds lie a bit less than two orders of magnitude above the SM predictions of Eq.~\eqref{eq:CR_SM_values}, $|\langle r^2_{\nu_{\ell\ell}}\rangle_{\rm
SM}| \lesssim 10^{-32}$~cm$^2$. Despite the larger exposure, they are only slightly stronger than those obtained from earlier XENONnT electron-recoil data~\cite{Giunti:2023yha}. They are also considerably weaker than the bounds obtained from the recent global analysis of laboratory \eves~and \cevns~data in Ref.~\cite{AtzoriCorona:2025xwr}, about one to two orders of magnitude for $\langle r^2_{\nu_{ee}}\rangle$ and a factor $\sim$5 for $\langle r^2_{\nu_{\tau\tau}}\rangle$. Next-generation detectors with larger exposures and lower backgrounds may significantly improve these sensitivities~\cite{Giunti:2023yha}.

Finally, as mentioned in Sec.~\ref{subsec:chradii}, the scattering of ultrarelativistic left-handed neutrinos cannot distinguish a charge radius from an anapole moment $a$. Our bounds can therefore be read as bounds on the corresponding anapole moments through $a_{\ell\ell'} = -\langle r^2_{\nu_{\ell\ell'}}\rangle/6$. At $1\sigma$ and without the $^{14}$C background, this gives, for instance, $a_{ee} \in [-4.1,\,18.5]\times10^{-32}$~cm$^2$ and $a_{\mu\mu/\tau\tau} \in [-18.1, \,17.0]\times10^{-32}$~cm$^2$. This interpretation is particularly relevant for Majorana neutrinos, whose
diagonal charge form factor vanishes identically, so that the anapole moment is their only diagonal electromagnetic
property~\cite{Giunti:2014ixa}.

\begin{figure}[t]
\centering
\includegraphics[width=\textwidth]{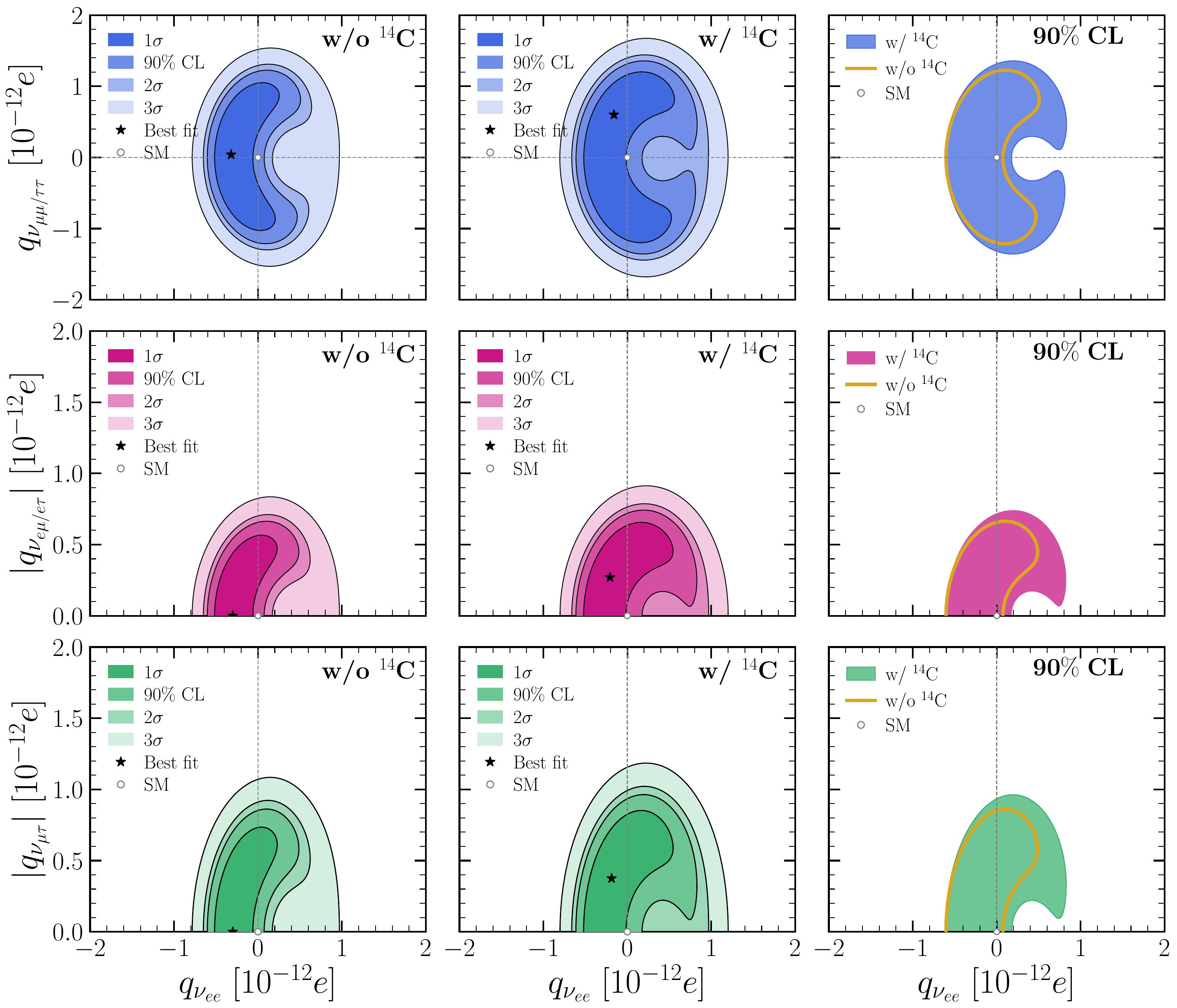}
\caption{Two-dimensional allowed regions for the neutrino electric charges, in the  $q_{\nu_{ee}}-q_{\nu_{\mu\mu/\tau\tau}}$ (top row), $q_{\nu_{ee}}-|q_{\nu_{e\mu/e\tau}}|$ (middle row), and $q_{\nu_{ee}}-|q_{\nu_{\mu\tau}}|$ (bottom row) planes. The left and middle columns show the results without and with the $^{14}$C background, respectively, at $1\sigma$, $90\%$ CL, $2\sigma$ and $3\sigma$; stars mark the best-fit points and the open circle the SM. The right column compares the $90\%$ CL regions with (filled) and without (orange contour) the $^{14}$C background.}
\label{fig:millicharge_2d}
\end{figure}

%%%%%%%%%%%%%%%%%%%%%%%%%%%%%%%%%%%%%%%%%%%%%%%%%%%%%%%%%
\subsection{Neutrino electric charges}
\label{subsec:res_MC}

\begin{figure}[!t]
\centering
\includegraphics[width=0.49\textwidth]{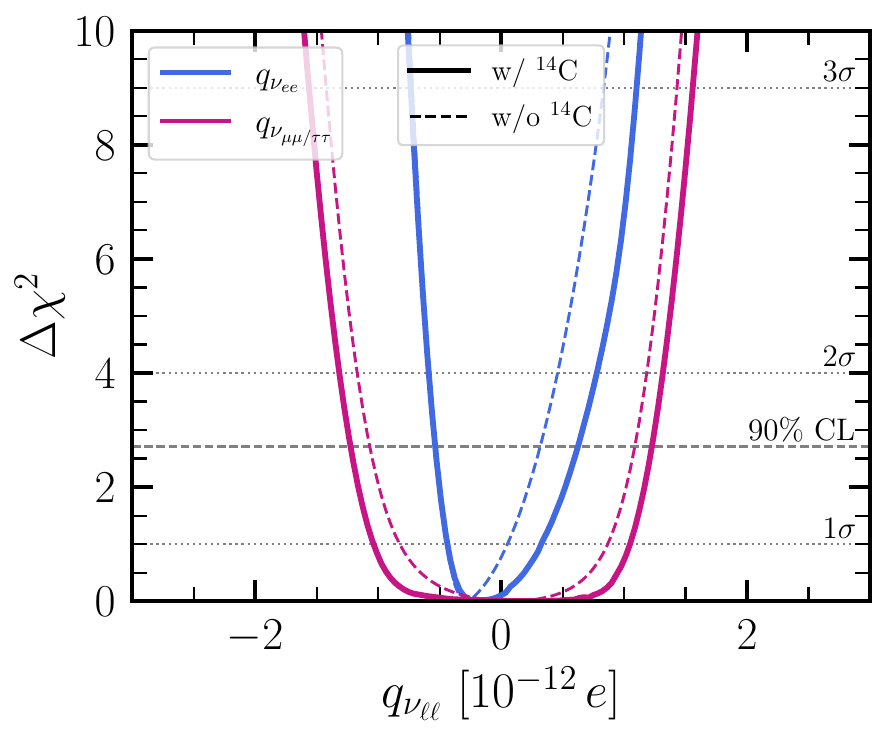}
\includegraphics[width=0.49\textwidth]{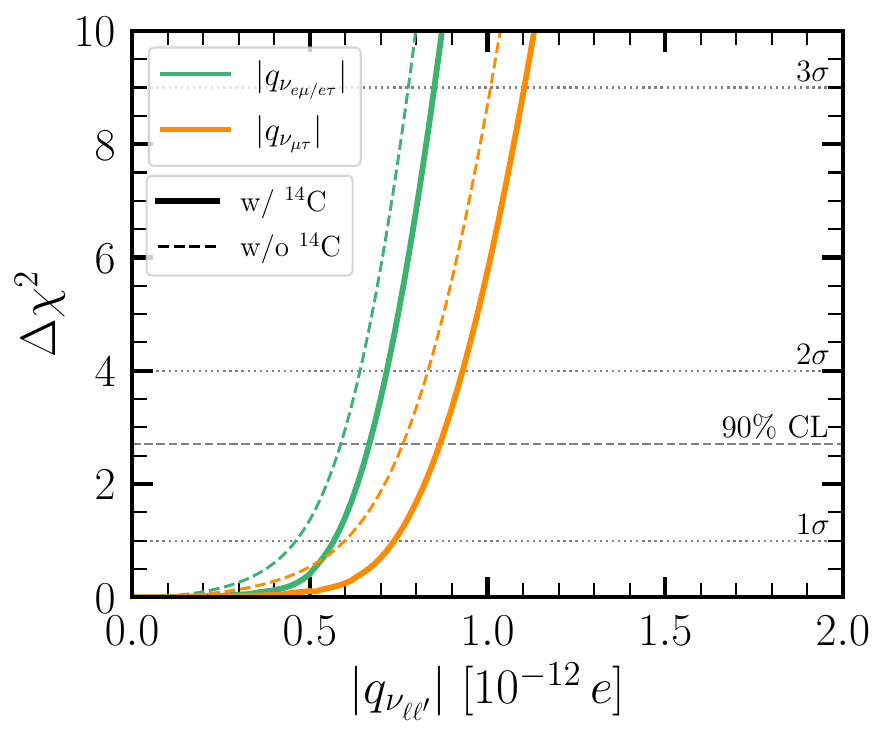}
\caption{Profiled $\Delta\chi^2$ for the diagonal (left) and transition (right) millicharges, with all other charges profiled over, with (solid) and without (dashed) the $^{14}$C background.}
\label{fig:millicharge_profiled}
\end{figure}

We next discuss neutrino electric charges. The allowed regions are shown in Fig.~\ref{fig:millicharge_2d} and the corresponding one-dimensional profiles in Fig.~\ref{fig:millicharge_profiled}, with all other charges profiled over in each case. As for the charge radii,
the solar flux at the detector does not allow us to separate the $\nu_\mu$ and $\nu_\tau$ components, and we present results for $q_{\nu_{ee}}$, $q_{\nu_{\mu\mu/\tau\tau}}$, $|q_{\nu_{e\mu/e\tau}}|$ and $|q_{\nu_{\mu\tau}}|$.
At $1\sigma$ we find
\begin{align}
q_{\nu_{ee}}^{\rm w/o\,{}^{14}C}
&=
\left(-0.30^{+0.35}_{-0.14}\right)\times10^{-12}\,e,
&
q_{\nu_{ee}}^{\rm w/\,{}^{14}C}
&=
\left(-0.19^{+0.51}_{-0.25}\right)\times10^{-12}\,e,
\nonumber\\
q_{\nu_{\mu\mu/\tau\tau}}^{\rm w/o\,{}^{14}C}
&=
\left(0.08^{+0.79}_{-0.90}\right)\times10^{-12}\,e,
&
q_{\nu_{\mu\mu/\tau\tau}}^{\rm w/\,{}^{14}C}
&=
\left(0.23^{+0.82}_{-1.26}\right)\times10^{-12}\,e,
\nonumber\\
|q_{\nu_{e\mu/e\tau}}|^{\rm w/o\,{}^{14}C}
&<
0.46\times10^{-12}\,e,
&
|q_{\nu_{e\mu/e\tau}}|^{\rm w/\,{}^{14}C}
&<
0.57\times10^{-12}\,e,
\nonumber\\
|q_{\nu_{\mu\tau}}|^{\rm w/o\,{}^{14}C}
&<
0.60\times10^{-12}\,e,
&
|q_{\nu_{\mu\tau}}|^{\rm w/\,{}^{14}C}
&<
0.74\times10^{-12}\,e.
\end{align}
The corresponding $90\%$ CL intervals are listed in Tab.~\ref{tab:em_limits}.

Unlike the charge radius, a millicharge leads to a characteristic spectral distortion. The diagonal charges enter through an energy-dependent shift of the vector coupling proportional to $1/T_e$, generating both an interference term with the SM amplitude and a purely electromagnetic contribution, with leading low-recoil behavior $1/T_e$ and $1/T_e^2$, respectively. Transition charges contribute incoherently, with a differential cross section exhibiting the same $1/T_e^2$ enhancement (see Sec.~\ref{subsec:millich}). Similar to the magnetic moment case, discussed below, the sensitivity is therefore dominated by the lowest-energy bins, and the $\mathcal{O}(1)$~keV threshold of XENONnT is the key to the strength of these bounds.

The interference with the SM amplitude determines the characteristic crescent shape of the allowed regions in Fig.~\ref{fig:millicharge_2d}. For $\nu_e$, with $g_V^e\simeq0.98$, the interference term is large:
negative values of $q_{\nu_{ee}}$ enhance the rate at low recoil energies and can accommodate the observed excess, which drives the best fit to $q_{\nu_{ee}}<0$, while small positive values suppress the rate and are disfavored. This suppression can be compensated by the incoherent contributions of the other charges, so positive values of $q_{\nu_{ee}}$ are allowed only when these are nonzero, and the allowed regions bend around an excluded region. The same effect makes the one-dimensional $q_{\nu_{ee}}$ profile in Fig.~\ref{fig:millicharge_profiled} strongly asymmetric. In the case of $\nu_{\mu,\tau}$, instead, $g_V^\mu\simeq-0.02$ is close to zero, the interference is negligible, and the $q_{\nu_{\mu\mu/\tau\tau}}$ profile is nearly symmetric, with a broader minimum.  The SM solution lies within the $1\sigma$ region in all cases.

As shown in the third column of Fig.~\ref{fig:millicharge_2d}, the inclusion of the $^{14}$C background has the following effect: as for the charge radii, part of the excess is absorbed by this background, so that rate-suppressing configurations are less disfavored. In turn, the region excluded at 90\% CL becomes smaller while the outer boundaries of the allowed $90$\% CL regions widen. This is reflected also in the 1D $\Delta \chi^2$ profiles of Fig.~\ref{fig:millicharge_profiled}.

Our bounds improve by about four orders of magnitude those obtained from a combined analysis of solar $^8$B \cevns~data, $|q_\nu|\sim\mathcal{O}(10^{-8})\,e$ at $90\%$ CL~\cite{DeRomeri:2024hvc}, due to the much lower recoil energies probed through \eves. They are comparable to those from earlier XENONnT, PandaX-4T and LZ electron-recoil data~\cite{Giunti:2023yha,A:2022acy,AtzoriCorona:2022jeb, Demirci:2025poc} and slightly stronger than those from reactor experiments, $|q_{\nu_e}| \lesssim 10^{-12}\,e$~\cite{Chen:2014dsa}. Much stronger, though more model-dependent, constraints arise from astrophysics, $|q_\nu| \lesssim 10^{-19} \,e$~\cite{Studenikin:2012vi}, and, assuming electric charge quantization, from the neutrality of matter, $|q_\nu|\lesssim 10^{-21}\,e$~\cite{Giunti:2014ixa}.

%%%%%%%%%%%%%%%%%%%%%%%%%%%%%%%%%%%%%%%%%%%%%%%%%%%%%%%%%
\subsection{Neutrino magnetic moments}
\label{subsec:res_MM}
% ============================================================
% MAGNETIC MOMENT
% ============================================================

\begin{figure}[t]
\centering
\includegraphics[width=0.55\textwidth]{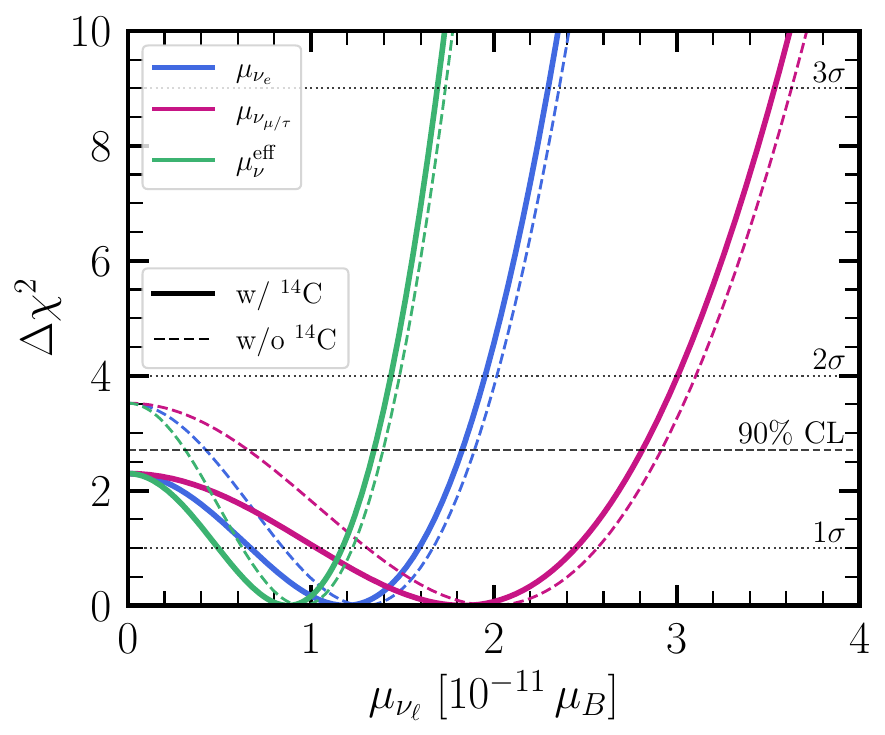}
\caption{Profiled $\Delta\chi^2$ for the neutrino magnetic moments $\mu_{\nu_e}$, $\mu_{\nu_\mu}$ (or, equivalently, $\mu_{\nu_\tau}$) and the flavor-universal $\mu_\nu^{\rm eff}$, with (solid) and without (dashed) the $^{14}$C background.}
\label{fig:magnetic_moment}
\end{figure}

The results of our analyses for effective magnetic moments are shown in Fig.~\ref{fig:magnetic_moment}. We consider three cases, switching on one parameter at a time: a magnetic moment of the electron neutrino only, $\mu_{\nu_e}$; of the muon neutrino only, $\mu_{\nu_\mu}$; and a flavor-universal moment, $\mu_\nu^{\rm eff} = \mu_{\nu_e} = \mu_{\nu_\mu} = \mu_{\nu_\tau}$. Since we assume $\sin^2\theta_{23} = 0.5$, the
$\nu_\mu$ and $\nu_\tau$ components of the solar flux at the detector are equal, and the results for $\mu_{\nu_\tau}$ coincide with those for $\mu_{\nu_\mu}$; we denote both by $\mu_{\nu_{\mu/\tau}}$. As discussed in Sec.~\ref{subsec:magmom}, each of these quantities receives contributions from the fundamental magnetic and electric dipole moments. At $1\sigma$ we obtain
\begin{align}
\label{eq:res_MM}
\mu_{\nu_e}^{\rm w/o\,{}^{14}C}
&=
\left(1.29^{+0.38}_{-0.44}\right)\times10^{-11}\,\mu_B,
&
\mu_{\nu_e}^{\rm w/\,{}^{14}C}
&=
\left(1.20^{+0.39}_{-0.53}\right)\times10^{-11}\,\mu_B,
\nonumber\\
\mu_{\nu_{\mu/\tau}}^{\rm w/o\,{}^{14}C}
&=
\left(1.99^{+0.58}_{-0.68}\right)\times10^{-11}\,\mu_B,
&
\mu_{\nu_{\mu/\tau}}^{\rm w/\,{}^{14}C}
&=
\left(1.84^{+0.61}_{-0.81}\right)\times10^{-11}\,\mu_B,
\nonumber\\
\mu_{\nu}^{\rm eff,\,w/o\,{}^{14}C}
&=
\left(0.95^{+0.28}_{-0.32}\right)\times10^{-11}\,\mu_B,
&
\mu_{\nu}^{\rm eff,\,w/\,{}^{14}C}
&=
\left(0.88^{+0.29}_{-0.39}\right)\times10^{-11}\,\mu_B.
\end{align}
In Tab.~\ref{tab:em_limits} we list also the corresponding 90\% CL intervals. The relative size of the three results follows
from the flavor composition of the solar flux at the detector: a moment of a single flavor is diluted by the fraction of that flavor in the flux, $P_{ee}\simeq 0.55$ for $\nu_e$ and $(1-P_{ee})/2\simeq0.23$ for each of $\nu_\mu$ and $\nu_\tau$.

In all cases the best fit lies at a nonzero magnetic moment. The no-magnetic-moment hypothesis corresponds to $\Delta\chi^2 \simeq 3.5$ ($\simeq1.9\sigma$) without and $\Delta\chi^2 \simeq 2.3$ ($\simeq1.5\sigma$) with the $^{14}$C background, identical for the three cases, since they differ only by the rescaling discussed above. Without $^{14}$C, a vanishing magnetic moment is therefore just outside the $90\%$ CL interval, whereas with $^{14}$C the $90\%$ CL results are
upper bounds. This preference is driven by the excess of events over the expected background discussed in Sec.~\ref{sec:res-SM}: the magnetic-moment contribution to the cross section grows as $1/T_e$ at low recoil energies, and can therefore accommodate an excess in the lowest-energy bins.
Unlike the charge radius or the weak mixing angle, which mainly change the overall rate, the magnetic moment has a distinct spectral shape, which also differs from the comparatively flat $^{14}$C spectrum at low energies. This explains why the inclusion of the $^{14}$C background shifts the magnetic-moment results by less than $10\%$. We stress that the preference for a nonzero magnetic moment is not statistically significant\footnote{A nonzero best fit is also found for the charge radii and millicharges. In those cases, however, several parameters are fitted simultaneously, and the same $\Delta\chi^2$ corresponds to a lower significance than in the one-parameter fits considered here.}.

Because of this preference for a finite effective magnetic moment, our upper bounds are slightly weaker than those obtained from earlier XENONnT, PandaX-4T, and LZ electron-recoil data~\cite{A:2022acy,AtzoriCorona:2022jeb,Khan:2022bel,Giunti:2023yha,XENON:2022ltv, Demirci:2025poc}.  On the other hand, they  improve upon those from the dedicated solar-neutrino experiment Borexino, $\mu_\nu^{\rm eff} < 2.8\times10^{-11}\,\mu_B$~\cite{Borexino:2017fbd} (see  Ref.~\cite{Coloma:2022umy} for the flavored bounds),
and from reactor experiments such as GEMMA, $\mu_{\nu_e} < 2.9\times10^{-11}\,\mu_B$~\cite{Beda:2012zz}, TEXONO $\mu_{\nu_e} < 7.4\times10^{-11}\,\mu_B$~\cite{TEXONO:2006xds} and CONUS~\cite{CONUS:2022qbb}, while astrophysical bounds from the red-giant branch are about an order of magnitude stronger, $\mu_\nu < 1.2\times10^{-12}\,\mu_B$~\cite{Capozzi:2020cbu} (for complementary astrophysical constraints from supernovae and
white-dwarf pulsations, see Refs.~\cite{Ayala:1998qz,Corsico:2014mpa},
respectively). The bounds here obtained are also about two orders of magnitude stronger than those obtained from solar $^8$B \cevns~data, $\mu_\nu^{\rm eff} < 1.1\times10^{-9}\,\mu_B$ at $90\%$ CL~\cite{DeRomeri:2024hvc}. Although the magnetic-moment contribution grows as $1/T$ in both channels, the SM cross section is much smaller for \eves~than for \cevns, where it is coherently enhanced by the squared weak charge of the nucleus. As a result, the magnetic-moment contribution starts to dominate over the SM one for much smaller values of $\mu_\nu$ in \eves. An overview of current bounds, including astrophysical ones, can be found in Ref.~\cite{Giunti:2024gec}. We note, however, that such comparisons should be taken with care: each experiment constrains its own effective magnetic moment, a combination of the fundamental moments that depends on the experimental environment~\cite{Miranda:2021kre,AristizabalSierra:2021fuc,Ternes:2025lqh}.

%%%%%%%%%%%%%%%%%%%%%%%%%%%%%%%%%%%%%%%%%%%%%%%%%%%%%%%%%%%%%%%%%%%%
% ============================================================
% STERILE DIPOLE PORTAL
% ============================================================

\subsection{Sterile dipole portal}
\label{subsec:res_DP}

\begin{figure}[t]
\centering
 \includegraphics[width=0.49\textwidth]{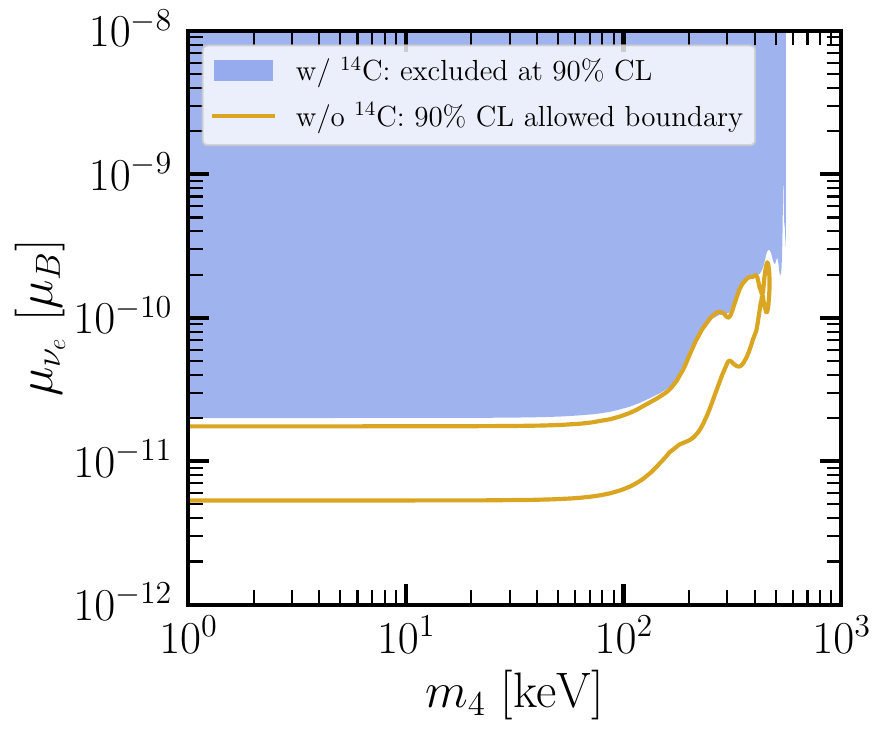}
 \includegraphics[width=0.49\textwidth]{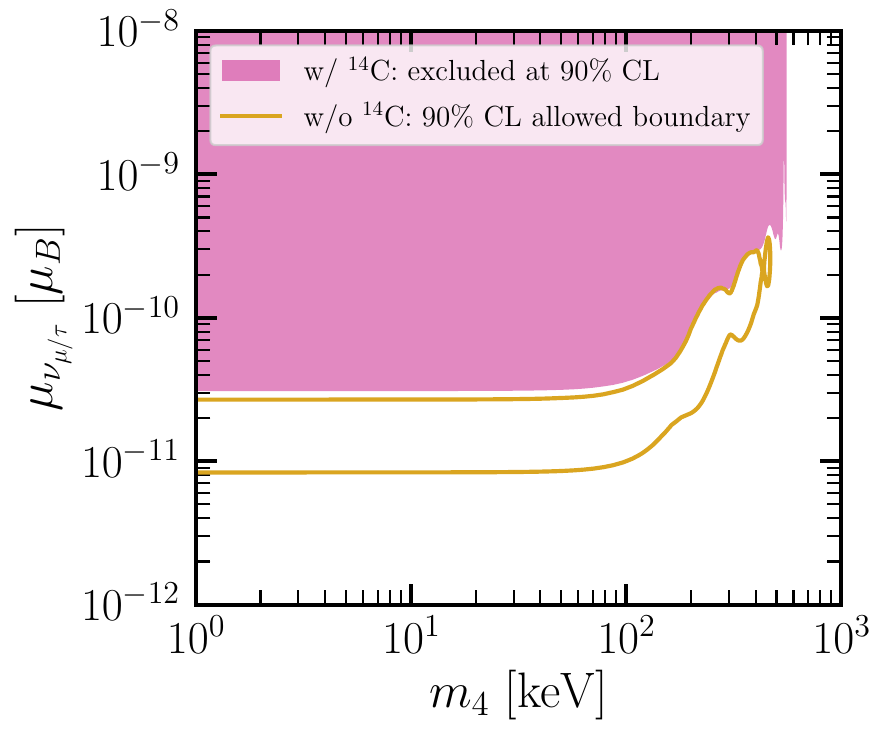}
 \caption{Constraints at $90\%$ CL on the sterile dipole portal in the $(m_4,\,\mu_{\nu_\ell})$ plane, for an active-sterile transition moment involving $\nu_e$ (left) and $\nu_\mu$ or $\nu_\tau$ (right). When the $^{14}$C background is included, the data yield an upper bound and the shaded region is excluded. When it is neglected, the data prefer a nonzero dipole coupling, and the allowed region is the band between the two orange curves.}
\label{fig:sterile_dipole}
\end{figure}

Our results for the sterile dipole portal are shown in Fig.~\ref{fig:sterile_dipole}, for a transition magnetic moment involving $\nu_e$ (left) and $\nu_\mu$ or $\nu_\tau$ (right), which, as in Sec.~\ref{subsec:res_MM}, give identical results. The light blue and pink shaded areas indicate the $90$\% CL exclusion contour, obtained when including the ${}{}^{14}$C background. When this background is not included the fit prefers at 90\% CL the regions inside the orange curves in both panels. 

For $m_4 \lesssim 100$~keV, the mass of the sterile state has a negligible impact on the kinematics, and the constraints reduce to those on the magnetic moment: with the $^{14}$C background we exclude $\mu_{\nu_e} \gtrsim 2\times10^{-11}\,\mu_B$ and $\mu_{\nu_{\mu/\tau}} \gtrsim 3\times10^{-11}\,\mu_B$. Without including the $^{14}$C background, instead, the same preference for a nonzero coupling found in Sec.~\ref{subsec:res_MM} appears as an allowed band depicted with orange contours. Since this preference is almost independent of $m_4$ in this mass range, the best-fit value of $m_4$ is not meaningful, and the best-fit coupling almost coincides with the magnetic-moment results of Eq.~\eqref{eq:res_MM}.

For heavier sterile states, the constraints weaken rapidly. Upscattering requires $E_\nu \geq m_4 + m_4^2/(2m_e)$ and, for a given $E_\nu$, the minimum recoil energy increases with $m_4$, removing the low-energy recoils where the dipole contribution is largest. Solar $pp$ neutrinos can produce $\nu_4$ only for $m_4 \lesssim 320$~keV, while the $^7$Be lines extend the reach to $m_4 \simeq 300$ and $560$~keV, respectively; these translate into the sharp kinematic cuts visible around those masses in Fig.~\ref{fig:sterile_dipole}.

These results improve on the bounds obtained from solar $^8$B \cevns~data~\cite{DeRomeri:2024hvc} by about two orders of magnitude for $m_4 \lesssim 0.5$~MeV, while remaining a factor of approximately $1.5$--$2$ weaker than the bounds from the combined analysis of XENONnT~\cite{XENON:2022ltv}, PandaX-4T~\cite{PandaX:2022ood}, and LZ~\cite{LZ:2022lsv} electron-recoil data presented in the same reference. That analysis yielded the strongest laboratory limits available at the time\footnote{An analysis of the updated PandaX-4T Run~0 and Run~1 electron-recoil datasets~\cite{PandaX:2024cic} was subsequently reported in Ref.~\cite{Mustamin:2025cpp},  resulting in slightly weaker bounds.} for $m_4 \lesssim 0.3$~MeV. We recall, however, that for such light sterile states the dipole couplings probed here are in tension with astrophysical and cosmological
observations, namely the cooling of SN1987A~\cite{Magill:2018jla,Chauhan:2024nfa}, Big Bang nucleosynthesis~\cite{Magill:2018jla,Brdar:2020quo}, and CMB
constraints on $\Delta N_\mathrm{eff}$~\cite{Brdar:2020quo}, although these bounds rely on additional assumptions about the production and decay of $\nu_4$ in these environments.

 %%%%%%%%%%%%%%%%%%%%%%%%%%%%%%%%%%%%%%%%%%%%%%%%%%%%%%%%%%%%%%%%%%%%
\section{Conclusions}
\label{sec:conclusions}
%%%%%%%%%%%%%%%%%%%%%%%%%%%%%%%%%%%%%%%%%%%%%%%%%%%%%%%%%%%%%%%%%%%%

\begin{table*}[t]
\centering
\begin{tabular}{lcc}
\toprule
Parameter
& Without $^{14}$C
& With $^{14}$C \\
\midrule
\multicolumn{3}{c}{\textit{Magnetic moments}} \\
\midrule
$\mu_{\nu_e}$
& $[0.43,\,1.90]\times10^{-11}\,\mu_B$
& $<1.83\times10^{-11}\,\mu_B$ \\

$\mu_{\nu_{\mu/\tau}}$
& $[0.66,\,2.92]\times10^{-11}\,\mu_B$
& $<2.81\times10^{-11}\,\mu_B$ \\

$\mu_{\nu}^{\rm eff}$
& $[0.31,\,1.40]\times10^{-11}\,\mu_B$
& $<1.35\times10^{-11}\,\mu_B$ \\

\midrule
\multicolumn{3}{c}{\textit{Neutrino millicharges}} \\
\midrule

$q_{\nu_{ee}}$
& $[-0.533,\,0.318]\times10^{-12}\,e$
& $[-0.535,\,0.625]\times10^{-12}\,e$ \\

$q_{\nu_{\mu\mu/\tau\tau}}$
& $[-1.067,\,1.088]\times10^{-12}\,e$
& $[-1.224,\,1.226]\times10^{-12}\,e$ \\

$|q_{\nu_{e\mu/e\tau}}|$
& $<0.587\times10^{-12}\,e$
& $<0.668\times10^{-12}\,e$ \\

$|q_{\nu_{\mu\tau}}|$
& $<0.762\times10^{-12}\,e$
& $<0.866\times10^{-12}\,e$ \\

\midrule
\multicolumn{3}{c}{\textit{Neutrino charge radii}} \\
\midrule

$\langle r^2_{\nu_{ee}}\rangle$
& $[-117.31,\,30.48]\times10^{-32}\,\mathrm{cm}^{2}$
& $[-114.76,\,28.05]\times10^{-32}\,\mathrm{cm}^{2}$ \\

$\langle r^2_{\nu_{\mu\mu/\tau\tau}}\rangle$
& $[-110.70,\,117.44]\times10^{-32}\,\mathrm{cm}^{2}$
& $[-107.01,\,113.62]\times10^{-32}\,\mathrm{cm}^{2}$ \\

$|\langle r^2_{\nu_{e\mu/e\tau}}\rangle|$
& $<62.40\times10^{-32}\,\mathrm{cm}^{2}$
& $<60.38\times10^{-32}\,\mathrm{cm}^{2}$ \\

$|\langle r^2_{\nu_{\mu\tau}}\rangle|$
& $<80.91\times10^{-32}\,\mathrm{cm}^{2}$
& $<78.29\times10^{-32}\,\mathrm{cm}^{2}$ \\

\bottomrule
\end{tabular}
\caption{One-dimensional $90\%$ CL allowed intervals for the neutrino electromagnetic parameters obtained from the XENONnT electron-recoil
data, without and with the $^{14}$C background. The magnetic moments are fitted one at a time; without $^{14}$C, the intervals exclude zero, reflecting the mild preference for a nonzero magnetic moment discussed in Sec.~\ref{subsec:res_MM}. For the millicharges and charge radii, the intervals are obtained by profiling over the remaining parameters of the same type. The background and solar-flux nuisance parameters are
profiled over in all cases.}
\label{tab:em_limits}
\end{table*}

The first observation of low-energy solar neutrinos through elastic neutrino-electron scattering in XENONnT, further supported by recent indications from PandaX-4T, opens a new chapter in the neutrino-physics program of dark matter direct detection facilities.  Thanks to low electron-recoil thresholds of $\mathcal{O}(1)$~keV, and a strong reduction of backgrounds, in particular from the $^{222}$Rn/$^{214}$Pb chain, these experiments are now sensitive to $pp$ and $^7$Be solar neutrinos, and can probe the neutrino-electron interaction at momentum transfers of $\mathcal{O}(0.1)$~MeV, in complementarity with dedicated laboratory probes. In this work, we have analyzed the new XENONnT data to extract a determination of the weak mixing angle at such low momentum transfer, and to constrain neutrino electromagnetic properties. Given the mild excess indicated by data, pointing to a $pp$ flux about $2\sigma$ above the SSM prediction, we have explored the role of a possible $^{14}$C background, whose spectrum closely mimics the solar-neutrino signal.

Within the SM, we have determined the $pp$ solar-neutrino flux. Without the $^{14}$C background, we find $\Phi_{pp} = (10.74^{+2.26}_{-2.25})\times10^{10}$~cm$^{-2}$\,s$^{-1}$, $2.1\sigma$ above the MB22m prediction, in line with the excess reported by the Collaboration. Including $^{14}$C reduces the flux to $(9.03^{+2.76}_{-2.41})\times10^{10}$~cm$^{-2}$\,s$^{-1}$ and the tension to $1.3\sigma$. The same excess, when the solar fluxes are fixed to their SSM values, drives the weak mixing angle at $\qtransfer\sim0.1$~MeV above its SM prediction, $\sin^2\theta_W = 0.465^{+0.075}_{-0.087}$ ($0.407^{+0.099}_{-0.115}$) without (with) $^{14}$C, corresponding to a $2.3\sigma$ ($1.4\sigma$) deviation. For the same reason, a meaningful determination of the electron-neutrino survival probability at $pp$ energies is not yet possible. Taken together with the agreement of other low-energy determinations of $\sin^2\theta_W$ with the SM, these results suggest that the excess is more likely related to the modeling of the low-energy backgrounds than to new physics.

Our bounds on neutrino electromagnetic properties are summarized in Tab.~\ref{tab:em_limits}. The magnetic moments are constrained at the
level of $\mu_\nu^{\rm eff} < 1.35\times10^{-11}\,\mu_B$ ($90\%$ CL, including the $^{14}$C background), about a factor of two stronger than the Borexino bound but slightly weaker than earlier XENONnT electron-recoil analyses, because of a mild preference for a nonzero magnetic moment induced by the current-data excess. The millicharges are constrained at the level of $|q_\nu|\lesssim10^{-12}\,e$, about four orders of magnitude better than solar $^8{\rm B}$ \cevns. The charge radii, which mainly affect the overall rate, are constrained to $\mathcal{O}(10^{-30})$~cm$^2$, somewhat less than two orders of magnitude above their SM values; these bounds apply equally to the anapole moments. Finally, for the sterile dipole portal we obtain bounds that almost coincide with the magnetic-moment ones for $m_4\lesssim100$~keV and extend up to $m_4\simeq560$~keV, set by the $^7$Be kinematic threshold; at low masses, however, this parameter space is in tension with astrophysical and cosmological observations.

The inclusion of the $^{14}$C background affects our results in different ways: it reduces the tensions in the solar fluxes and in $\sin^2\theta_W$, shrinks the regions excluded by the excess in the charge-radius and millicharge planes, and leaves the outer boundaries of the electromagnetic constraints almost unchanged. All in all, this shows that a precise characterization of the low-energy backgrounds, in particular of those whose spectra mimic the solar-neutrino signal, is essential for interpreting these data. 

Looking ahead, the precision of these measurements will improve substantially with larger exposures in XENONnT, LZ and PandaX-4T, and
with next-generation detectors such as the XLZD observatory. Clarifying the origin of the present excess will require a better characterization of the low-energy backgrounds, in particular a direct measurement or tighter constraint of the residual CO$_2$ contamination, as well as cross-checks with independent electron-recoil data from other detectors. Once this is achieved, these facilities will provide a competitive determination of the low-energy solar-neutrino fluxes and of $P_{ee}$ in the vacuum-dominated regime, an independent measurement of $\sin^2\theta_W$ at the lowest momentum transfers, and increasingly stringent tests of new physics in the neutrino sector, complementary to dedicated experiments.

\section*{Acknowledgments}
We are grateful to Riccardo Biondi for helpful discussions regarding backgrounds in XENONnT. V.D.R.~acknowledges financial support from CIDEXG/2022/20 (Generalitat Valenciana), and the Spanish grants PID2023-147306NB-I00 and CEX2023-001292-S (MCIU/AEI/10.13039/501100011033). D.K.P. acknowledges support from the European Union’s Horizon Europe research and innovation programme under the Marie Skłodowska‑Curie Actions grant agreement No.~101198541 (neutrinoSPHERE). D.K.P. benefits from the scientific environment of the Cluster of Excellence "Quantum Universe" funded by the Deutsche Forschungsgemeinschaft (DFG, German Research Foundation) under Germany’s Excellence Strategy (EXC 2121 “Quantum Universe”-390833306).

\appendix

\section{Validation of the solar neutrino flux analysis and impact of $\mathrm{^{14}C}$ background}
\label{app:flux_validation}

\begin{figure}[t]
\centering
\includegraphics[width=0.60\textwidth]{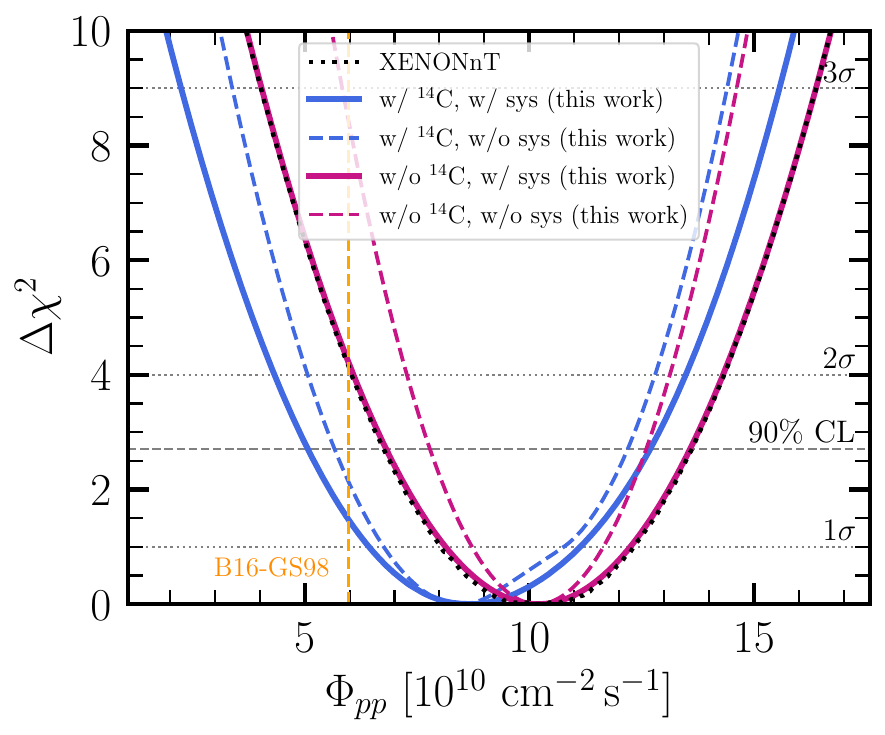}
\caption{Profiled $\Delta\chi^2$ for the $pp$ flux under the XENONnT assumptions (B16-GS98 solar model and a common normalization of the $pp$ and $^7$Be components), without (magenta) and with (blue) the
$^{14}$C background, the latter with the CO$_2$ concentration constrained to $[0,100]$~ppt in both SRs. Solid (dashed) curves include (neglect) the signal systematic uncertainty $\sigma_{\rm sys}$. The dotted curve shows the published XENONnT profile without $^{14}$C~\cite{XENON:2026rbz}, and the vertical (orange dashed) line the B16-GS98 prediction.}
\label{fig:flux_validation}
\end{figure}

In this appendix, we validate our analysis against the XENONnT one and assess the impact of our treatment of the $^{14}$C background.
To compare directly with the Collaboration's result, we first repeat the solar flux analysis under the same assumption as Ref.~\cite{XENON:2026rbz}, i.e., adopting the B16-GS98 solar model~\cite{Vinyoles:2016djt} and a common normalization for the $pp$ and $^{7}$Be components.
Without the $^{14}$C background, XENONnT reports, at $1\sigma$,
\begin{equation}
\Phi_{pp}^{\rm XENONnT}
=
(10.2\pm2.0)\times10^{10}\ {\rm cm}^{-2}\,{\rm s}^{-1}\, ,
\end{equation}
while we obtain (including the signal systematic uncertainty, see Sec.~\ref{sec:analysis})
\begin{equation}
\Phi_{pp}^{\rm w/o\,{}^{14}C}
=
\left(10.19^{+2.05}_{-2.06}\right)
\times10^{10}\ {\rm cm}^{-2}\,{\rm s}^{-1}\, ,
\end{equation}
is depicted as a solid magenta line in Fig.~\ref{fig:flux_validation}. 
When allowing for the presence of the additional background component $^{14}$C and constraining the concentration of CO$_2$, which is constrained to $[0,100]$~ppt in both SRs, the Collaboration finds, again at $1\sigma$,
\begin{equation}
\Phi_{pp}^{\rm XENONnT,\;w/\,{}^{14}C}
=
\left(8.7^{+2.8}_{-2.1}\right)
\times10^{10}\ {\rm cm}^{-2}\,{\rm s}^{-1}.
\end{equation}
Under the same assumptions, we find
\begin{equation}
\Phi_{pp}^{\rm w/\,{}^{14}C}
=
\left(8.64^{+2.49}_{-2.21}\right)
\times10^{10}\ {\rm cm}^{-2}\,{\rm s}^{-1},
\end{equation}
shown by the solid blue curve in Fig.~\ref{fig:flux_validation}. 
Both central values and uncertainties agree well with the XENONnT results, providing a solid validation of the robustness of our implementation of the signal, backgrounds and systematic uncertainties.
The dashed curves in Fig.~\ref{fig:flux_validation} show the results obtained when the signal systematic uncertainty $\sigma_{\rm sys}$ of Sec.~\ref{sec:analysis} is neglected: the profiles become significantly narrower, with the $1\sigma$ interval on $\Phi_{pp}$ reduced by about $29\%$ and $15\%$ without and with the $^{14}$C background, respectively, while their minima shift slightly toward larger fluxes. Including $\sigma_{\rm sys}$ is therefore necessary to reproduce the uncertainty quoted by the Collaboration.

\begin{figure}
    \centering
    \includegraphics[width=\linewidth]{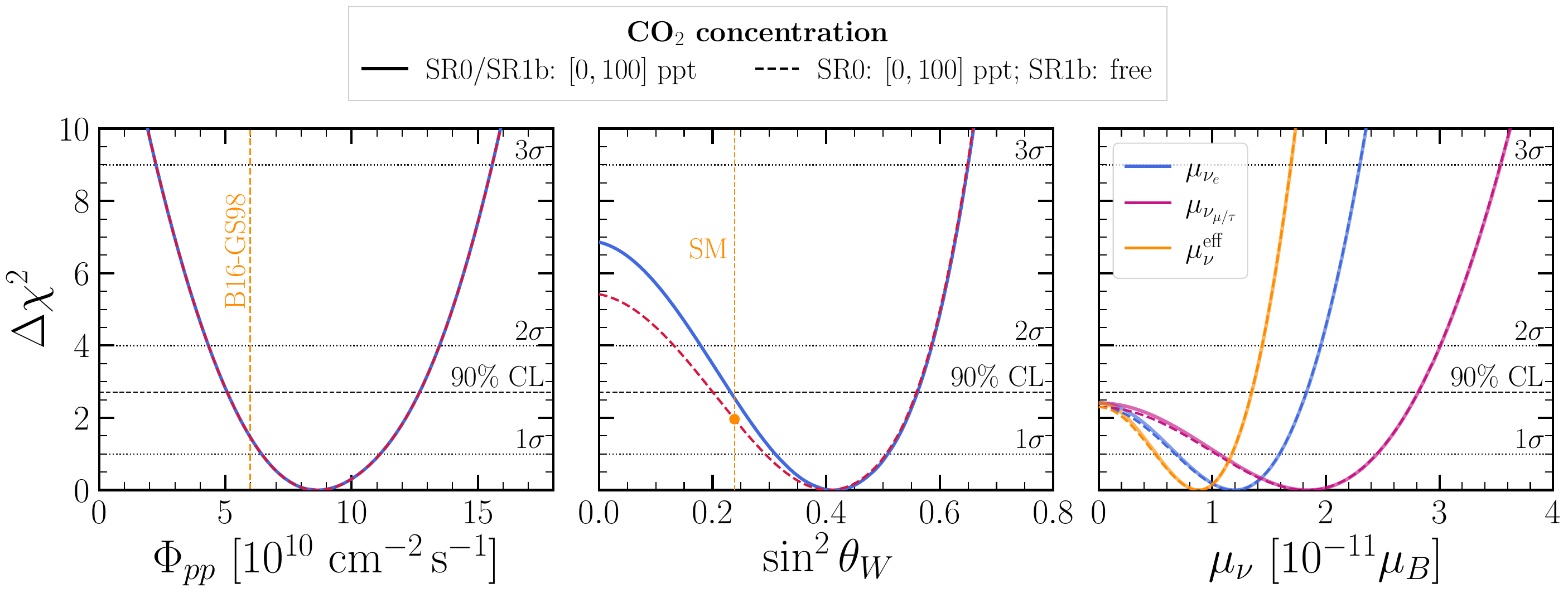}
    \caption{
    Profiled $\Delta\chi^2$ for two treatments of the $^{14}$C background. Solid curves: CO$_2$ concentration constrained to $[0,100]$~ppt independently in SR0 and SR1b, as in the XENONnT
    analysis. Dashed curves: our treatment, with SR0 constrained to $[0,100]$~ppt and the SR1b $^{14}$C normalization left free. From left to right: common $pp+{}^{7}$Be flux normalization (B16-GS98 inputs,
    expressed as $\Phi_{pp}$), weak mixing angle, and neutrino magnetic moments $\mu_{\nu_e}$, $\mu_{\nu_{\mu/\tau}}$ and $\mu_\nu^{\rm eff}$.
    The vertical lines mark the B16-GS98 and SM predictions; the dot marks the SM value of $\sin^2\theta_W$ on the profile for our treatment.}
    \label{fig:C14_concentration}
\end{figure}

We also explore the impact of our treatment of the $^{14}$C background. 
The XENON Collaboration constrains the CO$_2$ concentration to $[0,100]$~ppt in both SRs, based on the sensitivity of its gas purification system \cite{XENON:2026rbz}, whereas, as motivated in
Sec.~\ref{sec:analysis}, we leave the $^{14}$C normalization in SR1b free. 
We compare the two choices for three representative cases, which probe different features of the spectrum: the total ($pp+{}^7$Be) solar-flux normalization, which rescales the overall rate; the weak mixing angle, which changes the vector couplings and hence the relative
contributions of the different flavors; and the neutrino magnetic moments, which enhance the spectrum at low recoil energies.
As shown in Fig.~\ref{fig:C14_concentration}, leaving the SR1b concentration free has a negligible impact on the flux and magnetic-moment profiles.
A somewhat larger, though still moderate, dependence is observed for the $\sin^2 \theta_W$ case: the best fit and the upper side of the profile are also unchanged, but the lower side becomes shallower. In both cases, the SM value lies within the $90\%$ CL interval. Apart from this, our conclusions are not driven by the choice of the SR1b $^{14}$C treatment.

\bibliography{bibliography}  

\end{document}